\documentclass[aps,prd,twocolumn,preprintnumbers,nofootinbib,superscriptaddress,amsmath]{revtex4-2}
\usepackage{graphicx}
\usepackage{amsmath}
\usepackage{amssymb}
\usepackage{amsfonts}
\usepackage{hyperref}
\usepackage{url}
\usepackage{color}
\usepackage{bm}
\usepackage{array}
\usepackage{booktabs}
\usepackage[dvipsnames]{xcolor}
\usepackage{soul}
\usepackage{ragged2e}
\usepackage{subcaption}

\usepackage[normalem]{ulem}
\usepackage{xcolor}

\makeatother

\begin{document}

\preprint{IFT-UAM/CSIC-24-134}

\title{Astrometric constraints on stochastic gravitational wave background with neural networks}

\author{Marienza Caldarola}
\email{marienza.caldarola@csic.es}
\affiliation{Instituto de F\'isica Te\'orica UAM-CSIC, Universidad Aut\'onoma de Madrid, Cantoblanco, 28049 Madrid, Spain}

\author{Gonzalo Morrás}
\email{gonzalo.morras@uam.es}
\affiliation{Instituto de F\'isica Te\'orica UAM-CSIC, Universidad Aut\'onoma de Madrid, Cantoblanco, 28049 Madrid, Spain}

\author{Santiago Jaraba}
\email{santiago.jaraba-gomez@astro.unistra.fr}
\affiliation{Observatoire astronomique de Strasbourg, CNRS, Université de Strasbourg, 11 rue de l'Université, 67000 Strasbourg, France}

\author{Sachiko Kuroyanagi}
\email{sachiko.kuroyanagi@csic.es}
\affiliation{Instituto de F\'isica Te\'orica UAM-CSIC, Universidad Aut\'onoma de Madrid, Cantoblanco, 28049 Madrid, Spain}
\affiliation{Department of Physics and Astrophysics, Nagoya University, Nagoya, 464-8602, Japan}

\author{Savvas Nesseris}
\email{savvas.nesseris@csic.es}
\affiliation{Instituto de F\'isica Te\'orica UAM-CSIC, Universidad Aut\'onoma de Madrid, Cantoblanco, 28049 Madrid, Spain}

\author{Juan García-Bellido}
\email{juan.garciabellido@uam.es}
\affiliation{Instituto de F\'isica Te\'orica UAM-CSIC, Universidad Aut\'onoma de Madrid, Cantoblanco, 28049 Madrid, Spain}

\date{\today}

\begin{abstract}
Astrometric measurements provide a unique avenue for constraining the stochastic gravitational wave background (SGWB). In this work, we investigate the application of two neural network architectures, a fully connected network and a graph neural network, for analyzing astrometric data to detect the SGWB. Specifically, we generate mock Gaia astrometric measurements of the proper motions of sources and train two networks to predict the energy density of the SGWB, $\Omega_\text{GW}$. We evaluate the performance of both models under varying input datasets to assess their robustness across different configurations. We also perform a direct comparison with a likelihood-based approach using Markov chain Monte Carlo (MCMC) methods, finding out that the neural-network-based approach is significantly faster, taking on the order of minutes, compared to MCMC's order of days, while still capturing the same features in the data. Our results demonstrate that neural networks can effectively constrain the SGWB, showing promise as tools for addressing systematic uncertainties and modeling limitations that pose challenges for traditional likelihood-based methods.
\end{abstract}

\maketitle

% ---------------------------------------------------------
\section{Introduction}
\label{sec:intro}
The stochastic gravitational wave background (SGWB) arises from the superposition of numerous independent gravitational wave (GW) signals coming from all directions in the sky, produced by numerous independent sources throughout the Universe. These sources are categorized into cosmological and astrophysical origins, arising from various epochs in cosmic history. 

Cosmological sources of the SGWB include primordial processes such as inflation~\cite{Starobinsky:1979ty}, cosmic strings~\cite{Damour:2000wa}, and phase transitions~\cite{Kosowsky:1991ua} in the early Universe. In contrast, astrophysical sources contribute through events like compact binary mergers, supernovae, and rotating neutron stars~\cite{Regimbau:2011rp}.  The study of the SGWB is crucial, as it provides unique insights into the early Universe, fundamental physics, and astrophysical populations and processes. Detecting and characterizing the SGWB could validate inflationary models, probe the physics of the early Universe, and shed light on compact object populations across cosmic time~\cite{Caprini:2018mtu,Kuroyanagi:2018csn,Christensen_2018}.

Determining the origin of the SGWB is a challenging task requiring its characterization across a broad frequency range. Given the multitude of potential contributing sources, it is crucial to probe the SGWB amplitude at various frequencies. 
Future technological advancements and missions will probe a wide range of frequencies ranging from $10^{-3}$ to $10^8$Hz, from space-based observatories~\cite{Colpi:2024xhw,Kawamura:2020pcg} and ground-based detector networks~\cite{Abbott_2017} to ultrahigh-frequency GW experiments~\cite{Domcke:2023qle}, substantially improving sensitivity. In contrast, frequencies lower than $10^{-8}$Hz can be investigated using cosmological and astrophysical observations, such as the cosmic microwave background (CMB) B-mode polarization measurements~\cite{LiteBIRD:2020khw} and pulsar timing arrays (PTAs)~\cite{InternationalPulsarTimingArray:2023mzf}, with astrometry also playing a pivotal role.

Astrometry is the precise measurement of the positions and motions of celestial objects and offers a unique approach to probing the SGWB. GWs in the vicinity of Earth induce correlated distortions in the apparent positions and proper motions of distant sources. Therefore, the detection or non-detection of this coherent behavior in astrometric data enables the measurement or constraint of the SGWB~\cite{Pyne:1995iy,Jaffe:2004it,PhysRevD.83.024024,Mihaylov:2018uqm,Mihaylov:2019lft}.
By searching for quadrupole-correlated patterns in precise astrometric measurements, such as those from the Gaia mission~\cite{Gaia:2016zol}, it is possible to fill the gap in the frequency spectrum between CMB polarization and PTA measurements, enabling constraints on GWs in the $10^{-16} \text{ Hz} \lesssim f \lesssim 10^{-9} \text{ Hz}$ range. Astrometric constraints on the SGWB have been continuously updated over the past decades; see Refs.~\cite{Gwinn:1996gv,Titov:2010zn,Darling_2018,Jaraba:2023djs,Darling:2024myz}.

However, note that currently the PTA measurements~\cite{InternationalPulsarTimingArray:2023mzf} (around $10^{-9}$ Hz), the joint CMB+BBN constraint on the relativistic energy density~\cite{Yeh:2022heq} (for frequencies higher than $10^{-10}$ Hz), and the CMB $\mu$-distortion constraints from COBE/FIRAS~\cite{Kite:2020uix} (between $10^{-16}$ and $10^{-9}$ Hz) provide stronger limits on the SGWB amplitude. Nevertheless, with the anticipated high-precision proper motion measurements from the upcoming series of the Gaia data release, as well as the Nancy Grace Roman Space Telescope~\cite{Wang:2022sxn,Pardo:2023cag} and proposed upgrade mission THEIA~\cite{Malbet:2022lll,Garcia-Bellido:2021zgu}, astrometry has the potential to become a competitive tool for constraining the SGWB within this frequency window.

In this work, we explore the potential of neural networks (NNs) to analyze astrometric data and constrain the energy density of the SGWB. Traditional likelihood-based methods face several challenges in the analysis and do not fully leverage Gaia's astrometric measurements. First, due to the complexities in modeling the intrinsic proper motion of stars in our galaxy, current analyses are typically limited to distant quasars. While quasars offer the advantage of stability in their motion, their numbers are significantly smaller than the stars within our galaxy. Moreover, even with stable quasar catalogs, systematic errors in astrometric measurements and source misidentification remain major obstacles~\cite{Jaraba:2023djs}.

To fully realize the potential of astrometric surveys, it is crucial to increase the number of available sources by incorporating measurements of all stars in our galaxy. Achieving this goal entails overcoming several challenges: accurately modeling galactic rotation to subtract stars' intrinsic proper motions, establishing robust sample selection criteria, and addressing the substantial computational demands associated with analyzing billions of stars. Finally, future data releases will provide time-series measurements, offering significant advantages for constraining GWs~\cite{Moore:2017ity}. Additionally, synergy with PTA data presents an interesting avenue for further exploration~\cite{Qin:2018yhy,Caliskan:2023cqm,Inomata:2024kzr,Cruz:2024diu}. However, this will also substantially increase the complexity of the analysis. Addressing these issues will be critical to fully utilizing the power of astrometric data. 

The application of NNs to cosmological observations has rapidly developed across various experimental datasets, demonstrating significant advantages and driving revolutions in data analysis in the era of big data. The flexibility of NNs has the potential to overcome the difficulties and limitations we face with traditional methods. Our ultimate aim is to develop architectures capable of accurately constraining the SGWB using the vast amount of future astrometric data. As a first step, for the first time, we test the application of NNs to GW astrometry by using simulated Gaia quasar mock catalogs. In this preliminary study, we aim to lay the groundwork for future extensions and applications to real data.

In this work, we design two types of NNs that take astrometric data (positions, proper motions, etc.) as input and output a prediction of the SGWB amplitude, $\Omega_\text{GW}$. The first type is a fully connected network (FCN) known for its simplicity and flexibility. The second is a graph neural network (GNN) designed to process graph-structured data and capture relationships between elements. We place particular emphasis on the GNN due to the natural alignment of our dataset with a graph-based representation.

This paper is organized as follows. In Sec. \ref{sec:theory}, we provide an overview of the theoretical framework underlying GW astrometry and describe the details of simulating mock datasets. In Sec. \ref{sec:NNarchitecture}, we outline the architectures of the FCN and GNN used in this study. In Sec. \ref{sec:results}, we present the results of testing the performance of both architectures. First, we evaluate their performance using homogeneously distributed sources, followed by tests in inhomogeneous cases with galactic masks applied. Moreover, we perform a direct comparison with the standard Markov chain Monte Carlo (MCMC) methods. Finally, we conclude in Sec. \ref{sec:conclusions}.

% ---------------------------------------------------------
\section{Theoretical formalism and creation of mock data}
\label{sec:theory}
The amplitude of the SGWB is usually characterized by the energy density parameter~\cite{Maggiore:2007ulw,Romano:2016dpx}
\begin{equation}
    \Omega_\text{GW}(f) = \frac{1}{\rho_c} \frac{d\rho_\text{GW}}{d(\ln f)},
\end{equation}
where $\rho_{\rm GW}$ is the energy density of GWs and $\rho_c$ is the critical density for a flat universe,
\begin{equation}
    \rho_c = \frac{3H_0^2}{8\pi G},
\end{equation}
with $H_0 = 70 \, \text{km}\,\text{s}^{-1} \,\text{Mpc}^{-1} = 14.76\, \mu \text{as} \, \text{yr}^{-1}$ being the Hubble constant and $G$ being Newton's constant. 

In this paper, we want to address constraints on the SGWB coming from astrometric measurements, due to the deflection caused by GWs along the light trajectories of sources in the sky. It has been shown that the expected upper bound on the energy density $\Omega_\text{GW}$ is \cite{PhysRevD.83.024024}
\begin{equation}
    \label{eq:theor_Omega_gw}
    \Omega_\text{GW} \lesssim \frac{\Delta\mu^2}{NH_0^2},
\end{equation}
where $\Delta\mu^2$ is the variance associated with the proper motion at a given frequency, and $N$ the total number of sources. The magnitude of the proper motion $\mu$ is often decomposed as 
\begin{equation}
    \mu^2 = \mu_\delta^2 + \mu_\alpha^2\cos{\delta}^2,
\end{equation}
where $\mu_\delta$ represents the component of the proper motion in the direction of declination, while $\mu_\alpha$ denotes the component in the direction of right ascension.

To train the NNs and evaluate their performance through testing and validation, we create quasar mock catalogs using the Python package provided by the Gaia Collaboration, \textsc{pygaia}~\cite{pygaia}. Quasars are a suitable choice for the first step, as they have negligible intrinsic proper motion compared to other types of sources. Assuming a homogeneous distribution of sources, we randomly generate the right ascension and declination coordinates. The errors in the astrometric measurements depend on the brightness of the sources. Therefore, we randomly assign a G-band magnitude value~\cite{Storey-Fisher:2023gca,gaiadoc} for each source, drawn from a uniform distribution between $16 < G < 20.7$. The upper value corresponds to the nominal magnitude limit of Gaia, while the lower bound is determined by the quasar catalog presented in~\cite{Storey-Fisher:2023gca}. Then the \texttt{proper\_motion\_uncertainty} function from \texttt{pygaia.errors.astrometric} provides the associated errors for both the declination and right ascension components of each source, based on a given magnitude and the assumed Gaia data release. We assume the Gaia DR5 sensitivity. The details of the modeling for astrometric uncertainties can be found in the documentation provided by Gaia~\cite{gaiadoc}.

We then inject the SGWB signal, specifically the quadrupole component of the vector harmonics~\cite{Mignard:2012xm}. To achieve this, we randomly generate the values of the multipole coefficients of the vector spherical harmonics (see~\cite{Mignard:2012xm,Darling_2018} for the full expressions) from a uniform distribution over $[0, 1]$ and renormalize the amplitude using the relation Eq.~\eqref{eq:theor_Omega_gw}, which connects the quadrupole power $P_2$ to the SGWB amplitude $\Omega_\text{GW}$ as
\begin{equation}
    \label{eq:theor_Omega_gw_quadrupole}
    \Omega_\text{GW} \simeq \frac{6}{5}\frac{1}{4\pi} \frac{P_2}{H_0^2}.
\end{equation}
It is important to note that the SGWB also induces higher-order harmonics. However, our injection serves as a good approximation, as these higher-order contributions are subdominant relative to the quadrupole contribution~\cite{PhysRevD.83.024024}.

For training the NNs, the input set of parameters is characterized by six features: right ascension and declination coordinates, proper motion components for right ascension and declination, and the error amplitudes for right ascension and declination estimated from the magnitudes of the sources. We generate a set of mock catalogs, each with a different realization of noise and varying amplitudes of the SGWB injection, which are randomly selected from the linear-uniform distribution $[0,1]$. The NNs are trained to predict the correct values of $\Omega_\text{GW}$ injected based on the input astrometric measurements. We test three cases with different numbers of sources, chosen as $N_s = 500,\, 2000,\, 12000$. For each configuration with a fixed value of $N_s$, we generate 8000 mock simulated datasets and train the NNs independently for each configuration.

To demonstrate an example where NNs can showcase their flexibility, we consider an inhomogeneous distribution of sources. One potential cause of inhomogeneities in quasar samples is the masking of the galactic plane. To simulate such a case, we exclude sources in a symmetric central band around the celestial equator. Specifically, the declination coordinates of the sources are randomly and uniformly distributed above and below this equatorial exclusion zone, defined by the range $[-\delta, +\delta]$. We test four cases with $\delta = 10^{\circ}$, $20^{\circ}$, $40^{\circ}$, and $60^{\circ}$. 

% ---------------------------------------------------------
\section{Neural network architectures}
\label{sec:NNarchitecture}
The application of machine learning techniques to the analysis and interpretation of GW data represents a significant advancement in the field. These algorithms are particularly well suited for tackling complex problems, such as pattern recognition in large and intricate datasets, feature extraction, and significantly improving computation time.

The first NN we study is a FCN, where each neuron in a layer is connected to every neuron in the subsequent layer. It learns to detect patterns and features through a series of transformations and activation functions. Because of the density of connections, this kind of network is suitable to capture complex relationships in input data to have an estimate of the desired output. 
The final prediction is encoded in the output layer, which can represent different labels for classification problems or continuous values for regression problems. Specifically, in our case, we focus on a regression task aimed at extracting the value of $\Omega_\text{GW}$.

The second type of NN we focus on in this work is GNN~\cite{ZHOU202057,sanchez-lengeling2021a}, a deep NN developed to handle graph-structured data. A graph is a structure made by \textit{nodes} and relationships between nodes represented by connections called \textit{edges}, while information is encoded in features associated with nodes or edges and used for predictions. Nowadays, graph-structured data can be found extensively among various scientific and social disciplines, such as chemistry, social networks, and cosmology, making it important to deep learning models to handle such data. Consequently, interest in graph representation learning has grown significantly in recent years.

In this work, the NN architectures are implemented via the \small{\textsc{PyTorch}} framework. Specifically, we have carried out several tests to determine the best-performing architectures and hyperparameter settings for the networks, leveraging \textsc{Optuna} for optimization \cite{optuna_2019}. The training and validation datasets consisting of both the mock data and their corresponding target values are prepared by splitting the original dataset into 80\% for training and 20\% for validation \cite{sivakumar2024trade}.
% ---------------------------------------------------------
\begin{figure*}[t!]
\centering
\includegraphics[width=0.329\textwidth,trim = 0cm 0cm 0cm 0cm, clip]{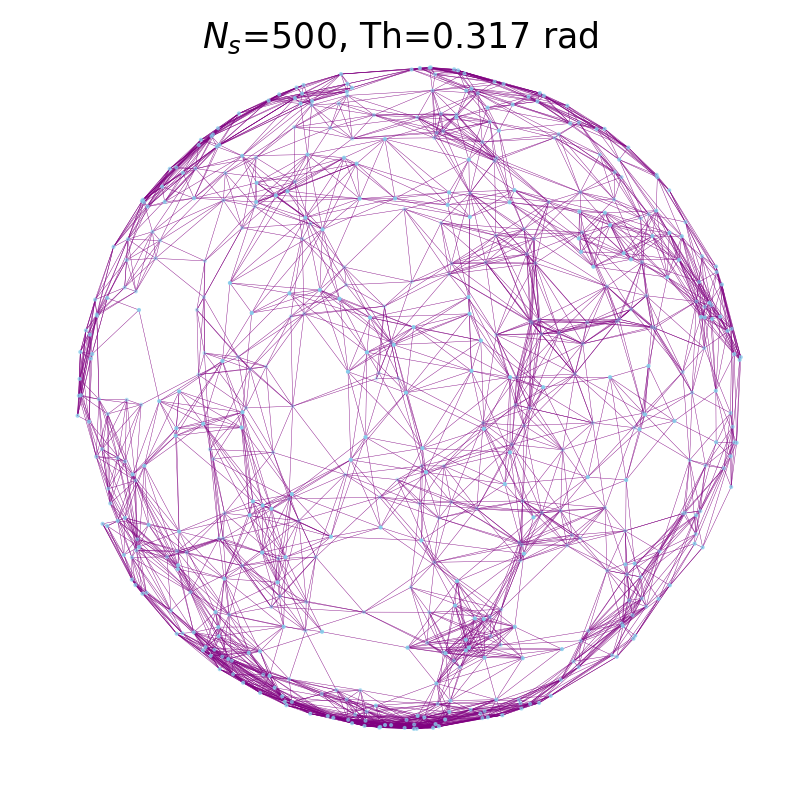}
\includegraphics[width=0.329\textwidth,trim = 0.07cm 0cm 0.07cm 0.0cm, clip]{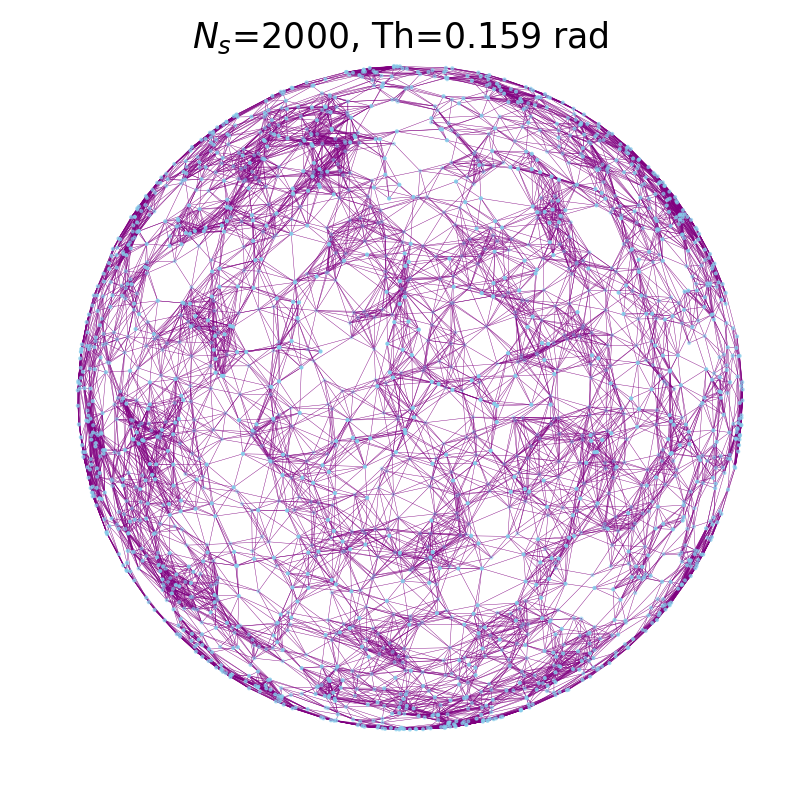}
\includegraphics[width=0.329\textwidth]{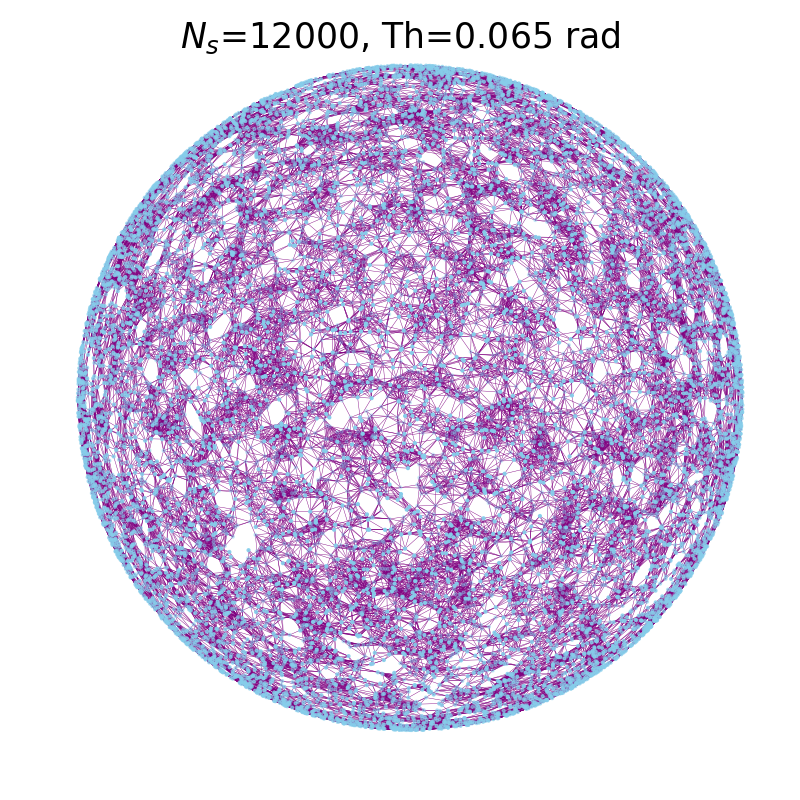}
\caption{\justifying Illustrative graph representation for the different number of sources. The number of edges for each graph has been chosen to ensure that almost all points are connected. In this plot, the corresponding number of edges for each configuration is $n_{\text{ed},500}=3062$, $n_{\text{ed},2000}=12758$, $n_{\text{ed},12000}=75746$, respectively.
\label{fig:graph3D}}
\end{figure*}

\subsubsection*{Fully connected network}
\label{subsec:FCN}
The input mock data are normalized by subtracting the mean and dividing by the standard deviation computed across the samples and their features, which helps to improve the stability and efficiency of the NNs training process \cite{lecun-98x}. The implemented FCN architecture consists of five hidden linear layers, each followed by a ReLU activation function~\cite{agarap2019deeplearningusingrectified}. The input size corresponds to the number of features (six in our case), and each hidden layer contains 256 neurons. After the final hidden layer, the output is averaged along the feature dimension to aggregate the learned representations. To reduce overfitting, dropout with a probability of 0.4 is applied. The resulting vector is then passed through a final linear layer that outputs a single scalar value, representing the network prediction for a regression task.

The loss function used is the MSELoss that measures the mean squared error (MSE) between each element in the prediction and true values, while the Adam optimizer~\cite{kingma2017adammethodstochasticoptimization} is used with a learning rate of $10^{-3}$. To regularize the training, we implement an early stopping technique to stop the training if the validation loss does not improve by a certain tolerance over 100 epochs. 

\subsubsection*{Graph neural network}
\label{subsec:GNN}
Among all the different types of GNN tasks, graph classification is used to classify entire graphs into various categories, based on structural graph features of a given dataset of graphs.

First, we create a set of graphs as input for the GNN. Each realization is converted into a graph, where nodes correspond to sources characterized by six astrophysical features. The normalization is the same as for the FCN training dataset. Edges between nodes are defined based on angular proximity. Specifically, we compute the Haversine distance between pairs of source positions and form an undirected edge if the distance falls below a chosen threshold. To ensure that each node is connected to its local neighborhood, the threshold is chosen slightly larger than the average distance, which scales as $\propto N_s^{-1/2}$, where $N_s$ is the number of sources. In our implementation, we use thresholds of $0.317$, $0.159$, and $0.065$ rad for $N_s=500$, $N_s=2000$, and $N_s=12000$, respectively. This procedure preserves the relevant local structure while avoiding unnecessary complexity.

To store the graph data, we use \texttt{Data} tool from \textsc{PyTorch Geometric} \cite{fey2019fast}, which contains the node features, edge connectivity, and a single graph-level label corresponding to the simulated value of $\Omega_{\text{GW}}$. The numbers of sources and realizations match those of the FCN. A three-dimensional visualization of example graphs for different $N_s$ values is shown in Fig.~\ref{fig:graph3D}.

The architecture consists of five graph convolutional layers implemented with \textsc{GCNConv}~\cite{kipf2017semisupervised}. Each convolutional layer is followed by a ReLU activation to introduce non-linearity. The input size corresponds to the number of node features, while each hidden layer contains 128 channels. After the final convolution, node embeddings are aggregated into graph-level representations using a global mean pooling operation, which averages over all nodes in a graph. To mitigate overfitting, a dropout layer with a probability of 0.4 is applied to the pooled representations. A final linear layer maps this representation to a single scalar output, yielding the predicted value of $\Omega_{\mathrm{GW}}$.
% ---------------------------------------------------------
\begin{figure*}[t!]
\centering
\includegraphics[width=0.48\textwidth,trim = 0cm 0cm 0cm 0cm, clip]{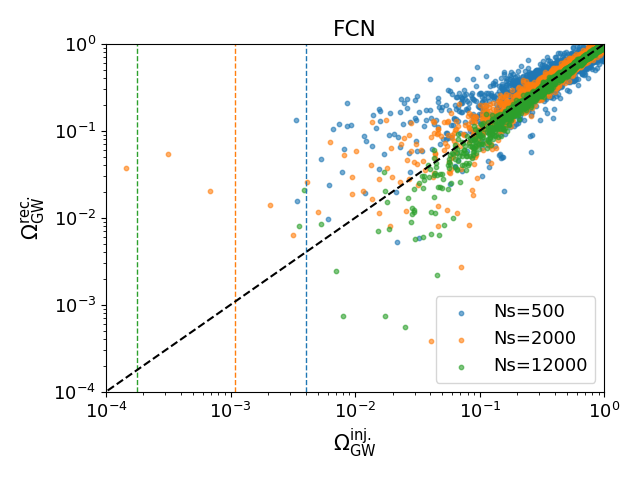}
\includegraphics[width=0.48\textwidth]{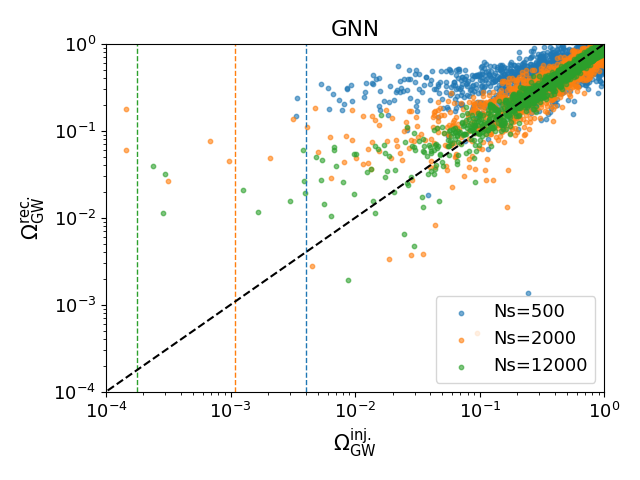}
\caption{\justifying Predictions in logarithmic scale on $\Omega_\text{GW}$ values, which were uniformly distributed in the range $[0,1]$. The left panel represents the result of FCN, while the right panel the one from GNN. The vertical line is the theoretical estimate for $\Omega_\text{GW}$ given by Eq.~(\ref{eq:theor_Omega_gw}). The predicted values are obtained using the test set ($1600$ samples, the $20\%$ of the original dataset of $8000$ mocks). 
\label{fig:nn_predictions}}
\end{figure*}

Training is performed using the Adam optimizer with a learning rate of $10^{-4}$ and MSE loss. Early stopping with a patience of 100 epochs and tolerance of $10^{-4}$ is employed to prevent overfitting.

% ---------------------------------------------------------
\section{Results \label{sec:results}}
\subsubsection*{Comparison between FCN and GNN}
In Fig.~\ref{fig:nn_predictions}, we present the test performance results for both FCN (left panel) and GNN (right panel). The networks can perform quite well for higher values of $\Omega_\text{GW}$, while the predictions start to become less precise for lower values when the noise starts to dominate.
The vertical line is the theoretically estimated sensitivity for $\Omega_\text{GW}$ given by Eq.~(\ref{eq:theor_Omega_gw}) included for reference. Note that this theoretical estimate assumes the same proper motion errors for all sources, whereas our mock data do not satisfy this assumption, as the proper motion errors depend on the magnitude of the sources. The results do not necessarily need to completely align with the theoretical line. As can be seen, by increasing the number of sources, the lines move toward the left, due to the inverse proportionality to the number of objects as seen in Eq.~(\ref{eq:theor_Omega_gw}) and the NNs predictions also improve with an increasing number of sources.

Note that the plot is shown in log scale, while our mock catalog dataset is generated with a linear-uniform distribution of $\Omega_\text{GW}$. We have tested both linear-uniform and log-uniform distribution of $\Omega_\text{GW}$ and found that NNs perform better with the linear-uniform distribution. Specifically, %the performance refers to 
for the linear-uniform case, we obtain 
a reduced variance in the estimator (tighter scatter) and a more consistent linear trend across the tested range of $\Omega_\text{GW}$. It is worth noting that the choice of training distribution represents an implicit prior, as NNs effectively learn the probability density of the training data. Consequently, the network's performance is intrinsically linked with these training settings, and the model may require retraining if the underlying physical prior for $\Omega_\text{GW}$ significantly deviates from our assumed distribution. Additionally, for the GNN, we evaluated the impact of varying the distance threshold values used to define the graph connectivity for a fixed number of sources. We observed that variations in the distance threshold did not lead to substantial differences in the trained performance of the GNN.

In both plots, the predictions follow the diagonal (black dashed line), showing that both networks recover the expected scaling of the signal. However, below $10^{-1}$, the predictions are more scattered. These differences can be linked to the intrinsic design of each architecture. Note that both architectures are already optimized, and further changes in complexity do not lead to better results, but only to a higher computational cost. In particular, for the GNN this could indicate that the scatter is not only a function of the model depth but it might also come from intrinsic factors such as how the network aggregates and propagates information within the graph structure.  Although deeper GNNs are better for capturing complex dependences, they also introduce challenges such as greater variability in predictions. These observations underscore the complex relationship between model design choices and performance outcomes, resulting in careful tuning of model architectures to balance precision and stability for specific tasks.
% ---------------------------------------------------------
\begin{figure*}[t!]
    \centering 
    \begin{subfigure}[b]{0.32\textwidth}
        \centering
        \includegraphics[width=\textwidth]{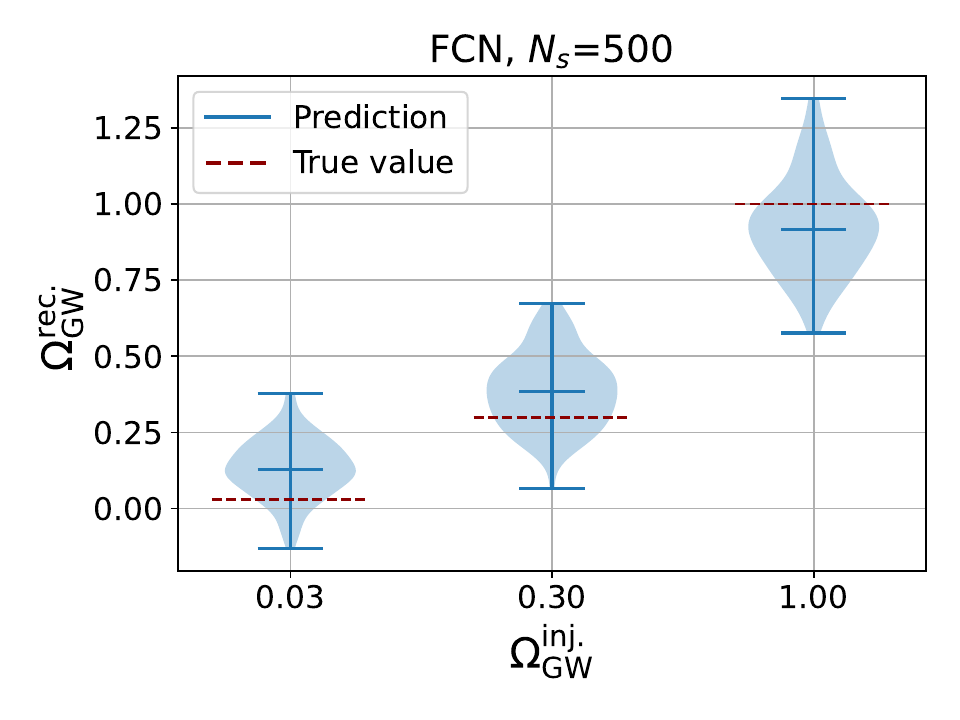}
    \end{subfigure}    
    \begin{subfigure}[b]{0.32\textwidth}
        \centering
        \includegraphics[width=\textwidth]{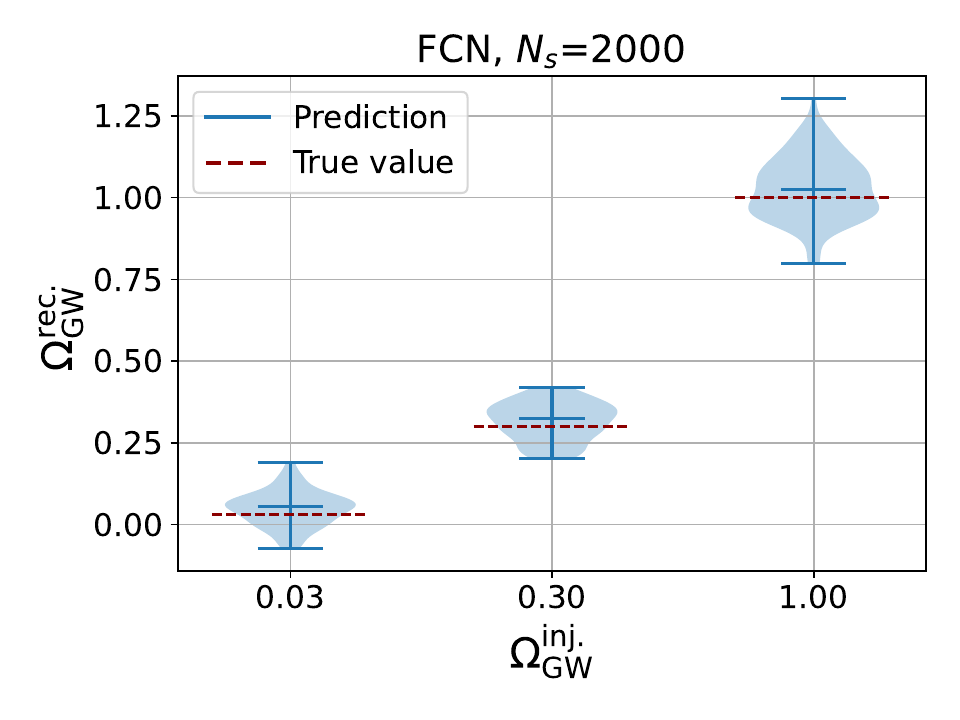}
    \end{subfigure}
    \begin{subfigure}[b]{0.32\textwidth}
        \centering
        \includegraphics[width=\textwidth]{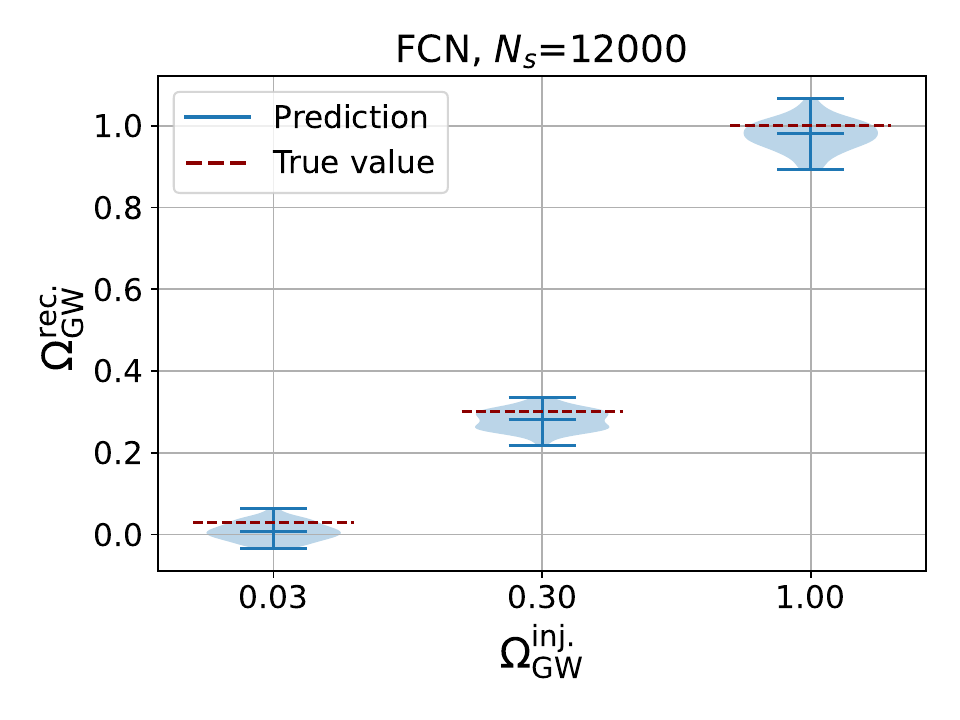}
    \end{subfigure}
    \begin{subfigure}[b]{0.32\textwidth}
        \centering
        \includegraphics[width=\textwidth]{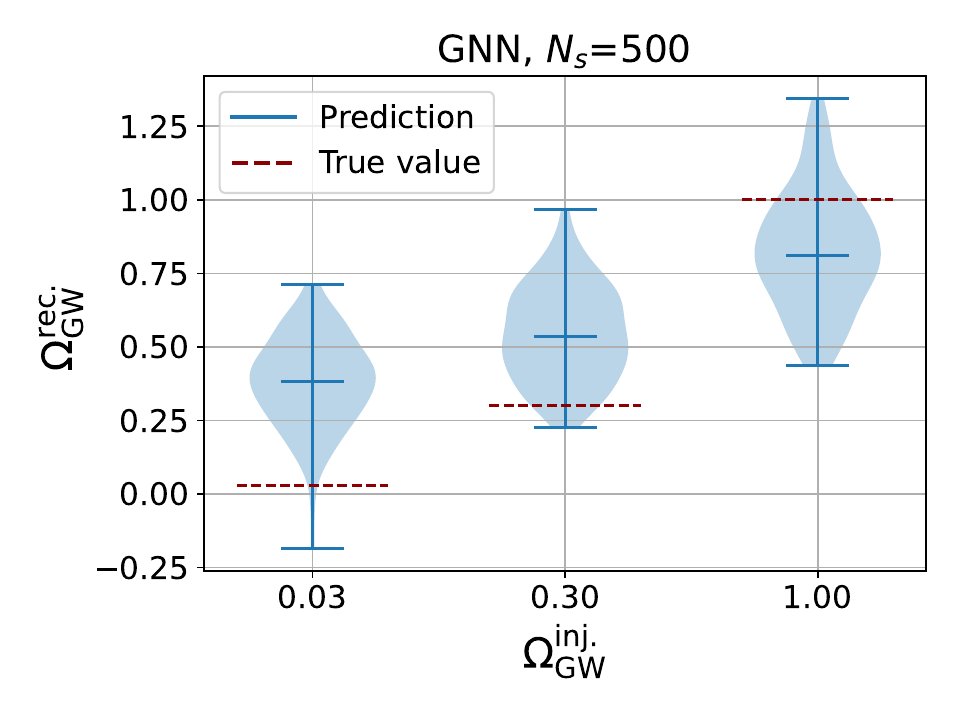}
    \end{subfigure}    
    \begin{subfigure}[b]{0.32\textwidth}
        \centering
        \includegraphics[width=\textwidth]{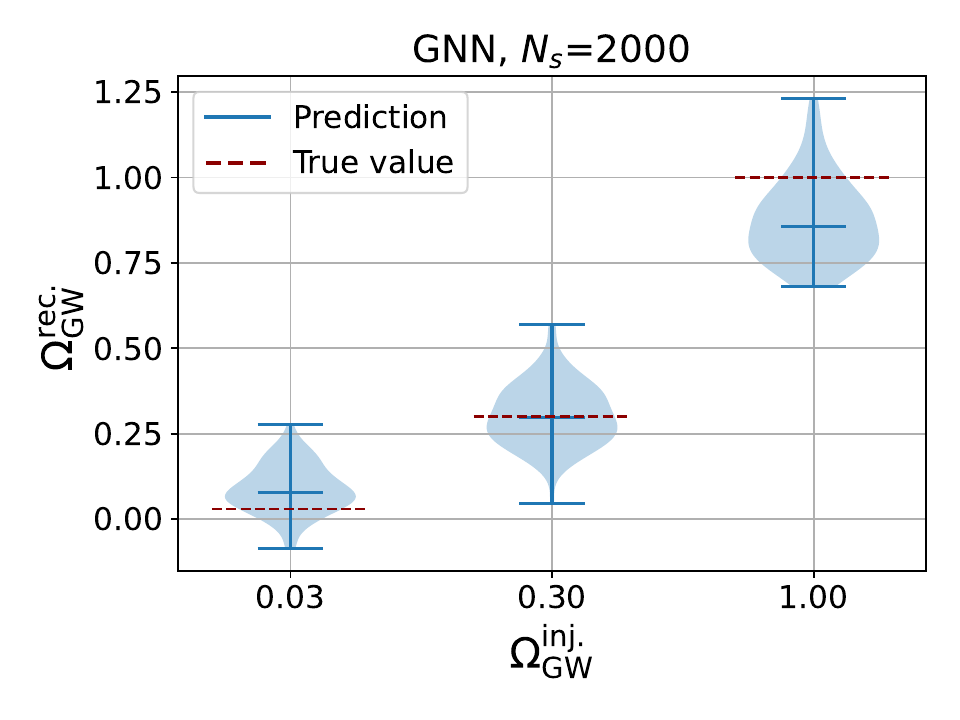}
    \end{subfigure}
    \begin{subfigure}[b]{0.32\textwidth}
        \centering
        \includegraphics[width=\textwidth]{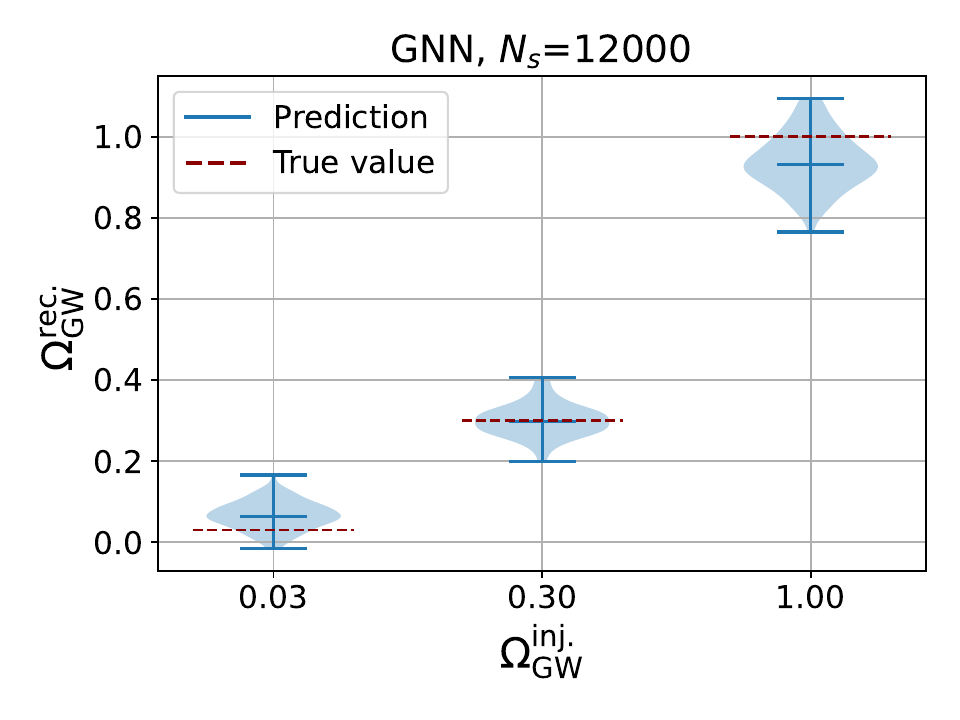}
    \end{subfigure} 
    \caption{\justifying Test of trained FCN and GNN networks to new test datasets obtained by fixing $\Omega_\text{GW}$ to $0.03\,,0.3\,,1.0$ values. Each violin illustrates the aggregated distribution of predictions across 100 independent realizations for each fixed value. %\delete{Comparison for different values of $N_s$. The corresponding MSE values are reported in Tab. \ref{tabel:mse}.}
    Results are shown for different values of $N_s$, with corresponding MSE values reported in Table \ref{tabel:mse}.}
    \label{fig:violin_plots}
\end{figure*}

Finally, to estimate the predictive uncertainties and assess generalization, we apply the trained NNs to new unseen datasets and compute the MSE between the predictions and the true values. This allows us to test how reliably the networks perform beyond the training data, which is essential for evaluating their robustness in practical applications. The new mock datasets are made up of $100$ new independent realizations by fixing the values of $\Omega_\text{GW}$ to $[0.03,\,0.3,\,1.0]$. %\delete{Results are reported in Fig. \ref{fig:violin_plots} and corresponding values in Tab. \ref{tabel:mse}.}
The violin plots in Fig.~\ref{fig:violin_plots} visualize the distribution of the predictions across these 100 realizations for each fixed value. The corresponding MSE values are reported in Table~\ref{tabel:mse}.

\setlength{\tabcolsep}{8pt}
\begin{table}[h]
\begin{center} {\footnotesize
\begin{tabular}{lccc}
\hline
& \multicolumn{3}{c}{MSE from FCN}  \\
$N_s$ & \multicolumn{1}{c}{$\Omega_\text{GW}=0.03$} & \multicolumn{1}{c}{$\Omega_\text{GW}=0.3$} & \multicolumn{1}{c}{$\Omega_\text{GW}=1.0$}\\
\hline
$500$   & $1.9626\times10^{-2}$ & $2.2690\times10^{-2}$ & $3.0275\times10^{-2}$ \\
$2000$  & $3.3112\times10^{-3}$ & $3.6179\times10^{-3}$ & $9.9824\times10^{-3}$ \\
$12000$  & $1.0028\times10^{-3}$ & $1.0326\times10^{-3}$ & $1.8717\times10^{-3}$ \\
\hline
& \multicolumn{3}{c}{MSE from GNN}  \\
$N_s$ & \multicolumn{1}{c}{$\Omega_\text{GW}=0.03$} & \multicolumn{1}{c}{$\Omega_\text{GW}=0.3$} & \multicolumn{1}{c}{$\Omega_\text{GW}=1.0$}\\
\hline
$500$   & $1.4777\times10^{-1}$ & $8.5298\times10^{-2}$ & $7.1031\times10^{-2}$ \\
$2000$  & $8.3903\times10^{-3}$ & $8.2136\times10^{-3}$ & $2.9764\times10^{-2}$ \\
$12000$  & $2.4736\times10^{-3}$ & $1.5479\times10^{-3}$ & $8.3370\times10^{-3}$ \\
\hline
\end{tabular} }
\end{center}
\caption{\footnotesize \justifying MSE values of predictions on the test dataset for both the FCN and the GNN. The comparison is shown across different prediction targets, specifically $\Omega_\text{GW}\in[0.03,0.3,1.0]$.}
\label{tabel:mse}
\end{table}

\subsubsection*{Comparison with MCMC}
\begin{figure}[t!]
\centering
\includegraphics[width=0.48\textwidth,trim = 0cm 0cm 0cm 0cm, clip]{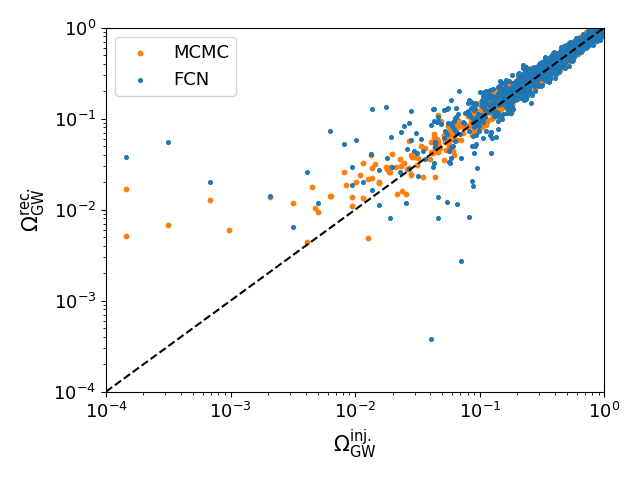}
\caption{\justifying Comparison in logarithmic scale between MCMC (orange) and FCN (blue) predictions on $\Omega_\text{GW}$ values, uniformly distributed in the range $[0,1]$. The number of test samples is $1600$ for the case $N_s=2000$. 
\label{fig:mcmc_vs_fcn_comparison}}
\end{figure}
It is of particular interest to compare the sensitivity of our NNs with that of likelihood-based methods, since MCMC approaches represent the state of the art in the literature \cite{Jaraba:2023djs}. In Fig. \ref{fig:mcmc_vs_fcn_comparison}, we show the comparison between the FCN and MCMC predictions across the entire original test dataset, which consists of $1600$ mock realization, for the representative case of $N_s=2000$. The MCMC algorithm used is the same as Ref.~\cite{Jaraba:2023djs}. We restrict the comparison with MCMC to the FCN case.

The MCMC model predictions appear tightly clustered around the diagonal, indicating high accuracy with potentially lower uncertainty. Nevertheless, the FCN seems to be performing close to the traditional estimator, generally following the same trend, displaying a slightly wider spread. Moreover, though FCN predictions exhibit a somewhat broader spread, they still capture the main trend accurately and crucially in a fraction of the time, generating predictions in order of minutes, compared to approximately the order of days required by the MCMC method for the same task. This highlights a major practical advantage of ML that is significantly faster inference, making them especially valuable when rapid predictions are needed or when working with large datasets. However, unlike the MCMC approach, which naturally provides full posterior distributions and credible intervals, the FCN approach introduced here provides point estimates only.

% ---------------------------------------------------------
\begin{figure*}[t!]
\includegraphics[width=1.0\textwidth]{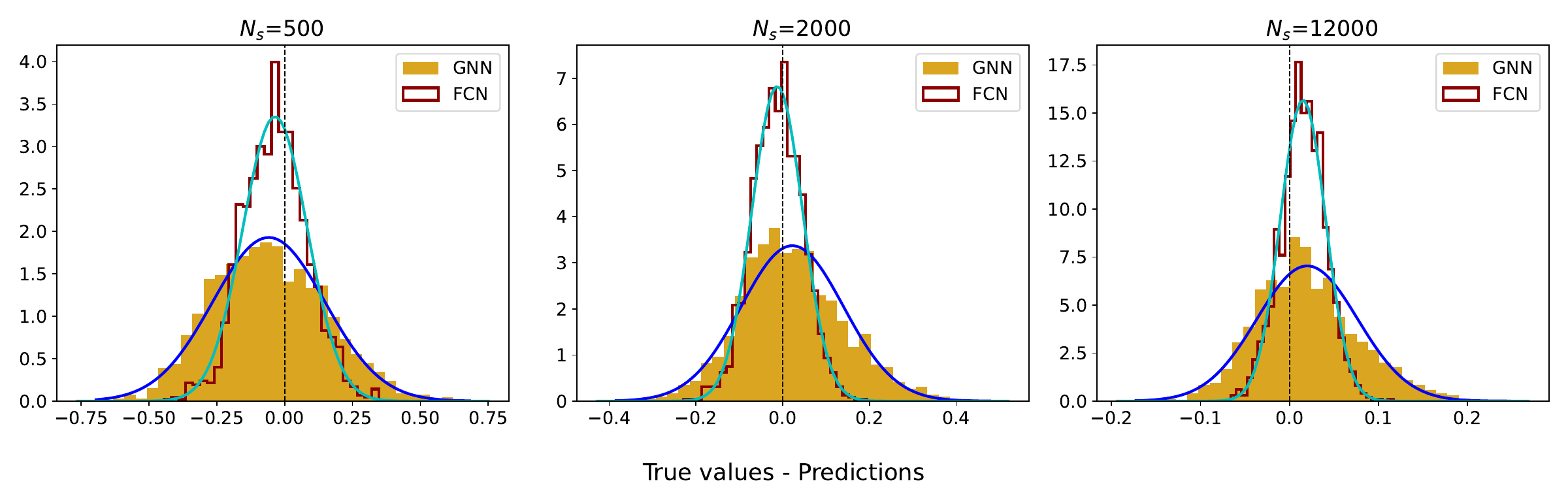}
\caption{\justifying The probability distribution of deviations between the true and predicted values for $\Omega_\text{GW}$ is plotted using the results obtained from the test dataset in Fig.~\ref{fig:nn_predictions}. The cyan curve represents the Gaussian fit to the FCN results, while the blue curve represents the Gaussian fit to the GNN results.  }
\label{fig:histo_fcn_vs_gnn}
\end{figure*}

% ---------------------------------------------------------
\begin{figure*}[t!]
\centering
\includegraphics[width=0.485\textwidth,trim = 0cm 0cm 0cm 0cm, clip]{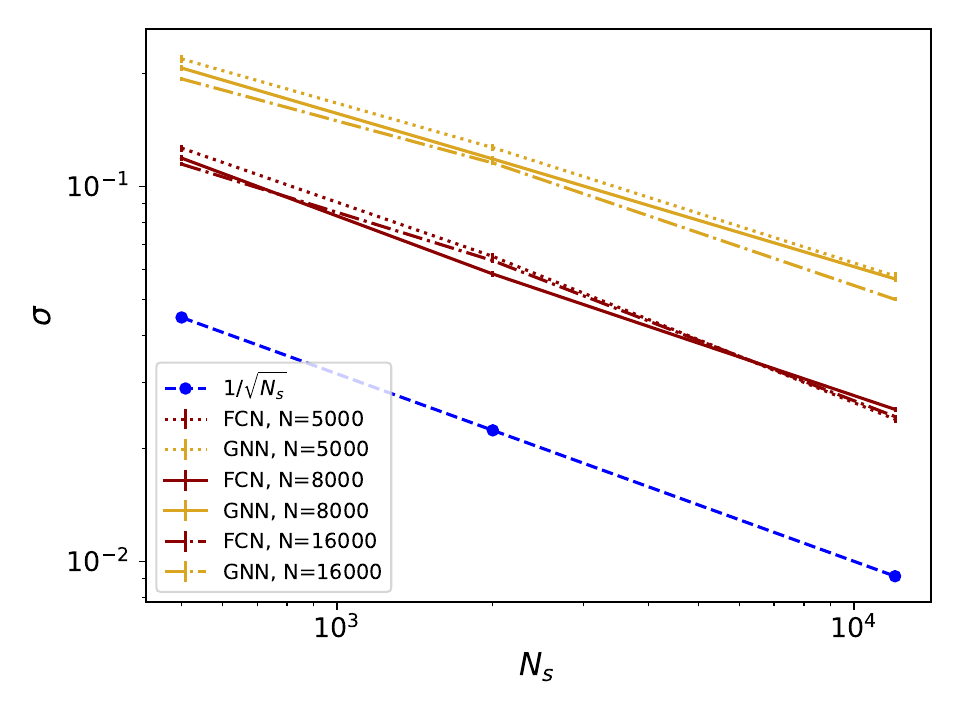}
\includegraphics[width=0.485\textwidth]{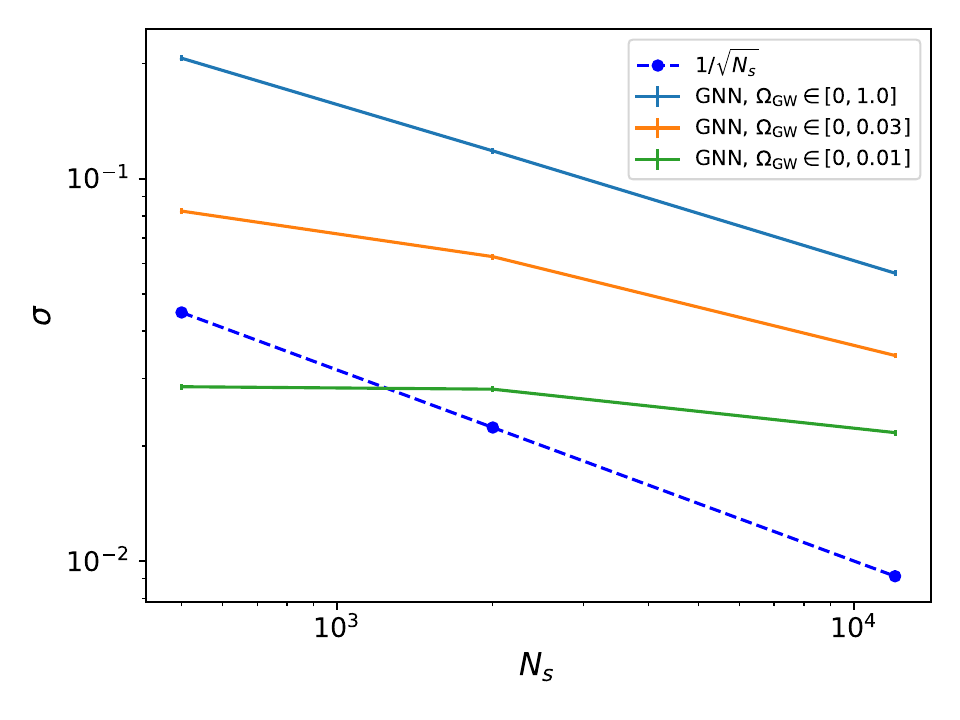}
\caption{\justifying The variance of the residual distributions in Fig.~\ref{fig:histo_fcn_vs_gnn} is plotted as a function of the number of sources. In the left panel, solid lines represent the results for the predictions in Fig.~\ref{fig:histo_fcn_vs_gnn} for the FCN (yellow) and GNN (brown). Other curves correspond to the results obtained by varying the number of mock data. These predictions are generated using test datasets, which constitute $20\%$ of the original mock datasets. The original mock datasets consist of $5000$ (dotted lines), $8000$ (solid lines), and $16000$ (dash-dotted lines) samples, respectively. The blue dashed line represents the theoretical sensitivity estimate, where the statistical error decreases proportionally to $1/\sqrt{N_s}$, with $N_s$ being the number of sources. In the right panel, the variance of the residual distributions is shown by varying the range of $\Omega_\text{GW}$ sampling. The default range of $[0, 1]$ is reduced to $[0, 0.3]$ and further to $[0, 0.1]$. The number of mock datasets is fixed at $8000$. Results are displayed only for the GNN.
\label{fig:width_histo_fcn_vs_gnn}
}
\end{figure*}
% ---------------------------------------------------------
\begin{figure*}[!t]
\centering
\includegraphics[width=0.48\textwidth,trim = 0cm 0cm 0cm 0cm, clip]{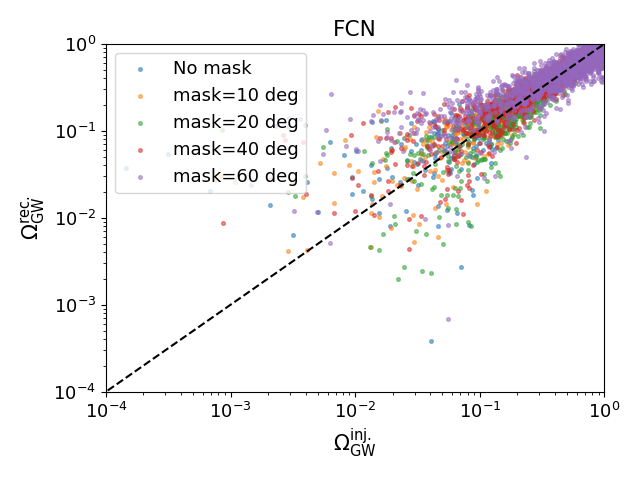}
\includegraphics[width=0.48\textwidth]{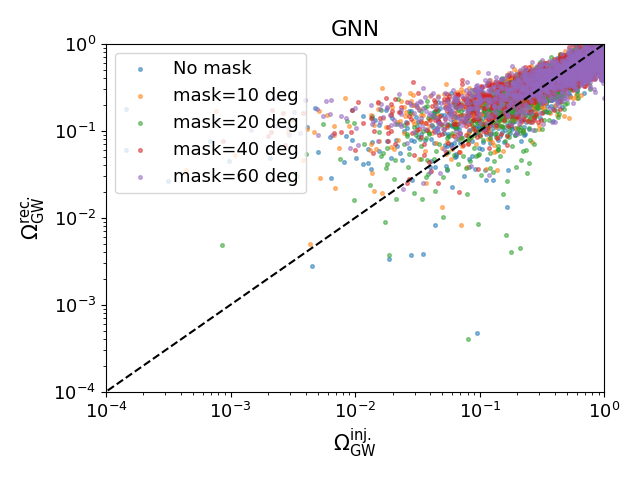}
\caption{\justifying Predicted values of $\Omega_\text{GW}$ for the masked dataset are tested by changing the area of the mask. The number of sources is fixed to $N_s=2000$. The left panel represents the result of FCN, while the right panel the one from the GNN. \label{fig:nn_predictions_mask}}
\end{figure*}
% ---------------------------------------------------------
\begin{figure*}[!t]
\includegraphics[width=1\textwidth]{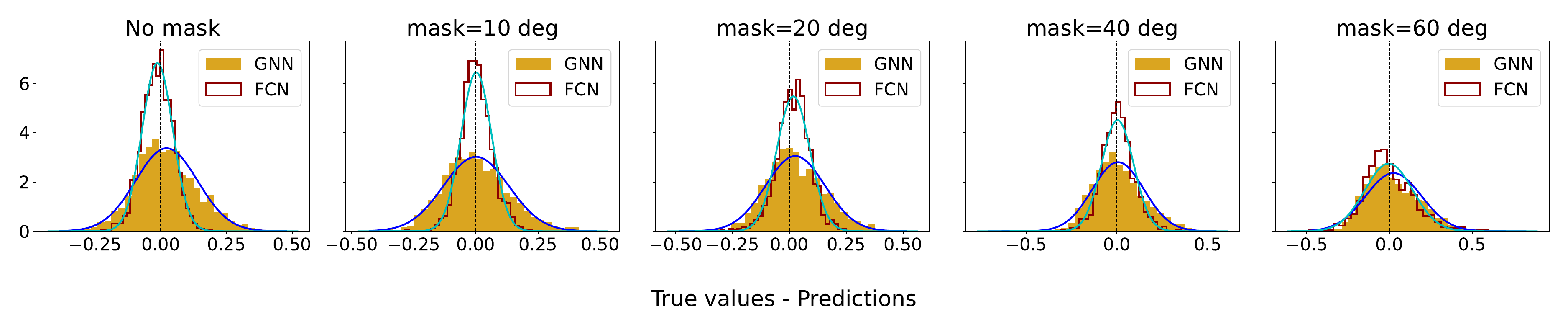}
\caption{\justifying The probability distributions of deviations between the true and predicted values for $\Omega_\text{GW}$ are plotted using the results obtained from the test dataset in Fig.~\ref{fig:nn_predictions_mask}, in the case of masked datasets. The cyan curve represents the Gaussian fit to the FCN results, while the blue curve represents the Gaussian fit to the GNN results.
\label{fig:histo_fcn_vs_gnn_mask}}
\end{figure*}

\subsubsection*{Distributions of the variance and its dependence on the source number}
To further analyze the performance of the models, we plot in Fig.~\ref{fig:histo_fcn_vs_gnn} the distribution of the residuals, i.e., the difference between the predicted and true values. We observe that the variance of the FCN is generally smaller than that of the GNN. For reference, we also include the Gaussian fit of the histogram in the plot.

Furthermore, we observe a clear trend where the width of the distribution decreases as the number of sources increases. To investigate this trend more explicitly, Fig.\ref{fig:width_histo_fcn_vs_gnn} shows the variance as a function of the number of sources $N_s$. For reference, we also plot a line proportional to $N_s^{-1/2}$.

The left panel compares the FCN and GNN results. Consistent with Fig.~\ref{fig:histo_fcn_vs_gnn}, the FCN exhibits a smaller variance compared to the GNN. To investigate this further, we conduct additional tests by varying the number of mock data $N$ plotted in the same figure based on the hypothesis that the requirement for more catalog realizations becomes increasingly critical as $N_s$ grows. We observe that the result is not affected when comparing the lines for $N=5000$ and $N=8000$.

In the right panel, we show that the range of $\Omega_\text{GW}$ for the mock data should be optimized based on the sensitivity.

\subsubsection*{Inhomogeneous case: masked samples}
Above, we have demonstrated the application of NNs with simplified mock data settings. However, there are many complex factors to consider when handling real data, and this presents a much longer path to fully realizing the potential of NNs. We leave it for the future work, while here, to illustrate one example of the flexibility of NNs, we apply them to masked data and evaluate their performance.

We fix the number of sources to $N_s=2000$ and mask the regions of the dataset corresponding to the declination of $\delta=10^{\circ}$, $20^{\circ}$, $40^{\circ}$, and $60^{\circ}$. While we are aware that $\delta = 60^{\circ}$ is a non-realistic value for the mask, it is used here solely for training performance purposes. The same NN architectures are trained on these masked subsets and the results are shown in Figs.~\ref{fig:nn_predictions_mask} and \ref{fig:histo_fcn_vs_gnn_mask}.  

Despite the variation in input data coverage, the overall behavior of the two architectures remains qualitatively consistent, as seen in Fig.~\ref{fig:nn_predictions_mask}. These results suggest that both networks remain robust under inhomogeneous sky coverage, with performance gradually degrading as masking increases. This behavior is evident in the distributions shown in Fig.~\ref{fig:histo_fcn_vs_gnn_mask}, where the variance clearly rises with the mask size. The FCN tends to produce slightly narrower residual distributions, indicating marginally better performance in terms of accuracy, whereas the GNN displays a broader spread around the true values. However, one can notice from Fig.~\ref{fig:histo_fcn_vs_gnn_mask} that as the mask size increases, the FCN performance degrades more noticeably than that of the GNN. This suggests that the GNN, while less accurate in the absence of strong masking, is comparatively more robust to missing or more complicated datasets. Both networks successfully recover the dominant $\Omega_{\rm GW}$ signal, but the dispersion of predictions increases systematically as the masked region grows, in agreement with expectations.

\subsubsection*{Computation time}
\begin{figure}[t!]
\vspace{-0.5cm}
\includegraphics[width=0.5\textwidth]{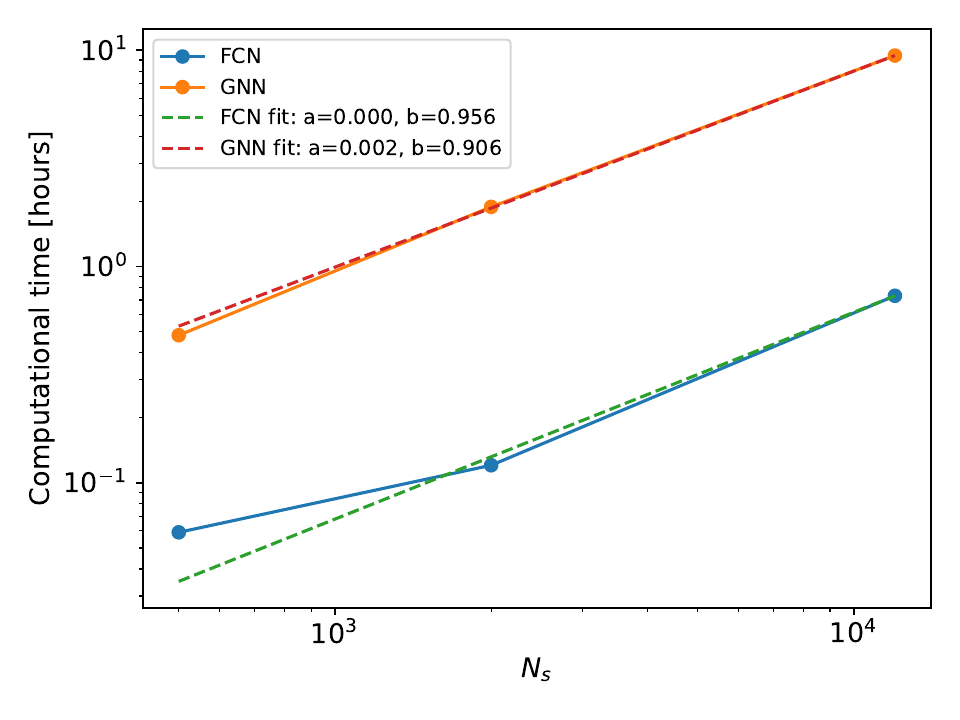}
\caption{\justifying Plot of the computational time in a logarithmic scale. The dashed lines correspond to a power-law fit of the form $y(x)=a\cdot x^b$. The best-fit values for $a$ and $b$ are shown in the legend. \label{fig:comp_time}}
\end{figure}

Given the expected number of sources available in future data, a reduction in computation time is primarily anticipated for the NNs. To address this, the computational time (including the training time) is plotted in Fig.~\ref{fig:comp_time} as a function of the number of sources. All these results were evaluated on a GPU cluster.\footnote{For completeness, the specs of the GPU cluster are an INTEL ICE LAKE 6338N 2.20GHZ CPU, with 32 GB DDR4 RAM, and four NVIDIA A100 80 GB CoWoS HBM2 PCIe 4.0 GPUs, along with three HDDs of 4 TB each, and an Infiniband adapter for fast connectivity between the nodes.} As observed, in both cases, the computation time increases logarithmically with the number of sources.
The GNN exhibits higher computational
time, which highlights the greater computational complexity inherent to the network architecture and the handling of more complex input data. In both cases, the behavior reflects a more accentuated computational cost for larger datasets.

Both networks are competitive in predicting the value of $\Omega_\text{GW}$, while the FCN shows an advantage in terms of the computation time. This is probably because of the architectures, and the GNN requires handling more complex graph structures. This complexity may make the implementation and computation less efficient and less competitive.
% ---------------------------------------------------------

% ---------------------------------------------------------
\section{Conclusions}
\label{sec:conclusions}
In this work, we have explored the possibility of constraining SGWB through astrometric measurements, with a particular focus on the application of NNs. These networks could provide complementary tools to traditional methods for detecting and constraining the SGWB, which is an interesting possibility, especially in the context of current or upcoming surveys that could uncover new aspects of the cosmological history of the Universe.

Specifically, we analyze the performance of two NN architectures, the FCN and GNN on a regression task to predict the value of the SGWB amplitude $\Omega_\text{GW}$. The results highlight distinct strengths and limitations for each architecture.

It is worth noting that the FCN is more reliable when a complete (original) dataset is available, as it exhibits a higher degree of precision in the prediction of the injected values, showing tighter clustering around the diagonal compared to those of the GNN, particularly as the number of samples increases. The corresponding histograms of the distribution of injection-predictions reinforce this, with the FCN distributions being narrower and centered closer to zero, indicating smaller prediction errors. This suggests that the FCN, when adequately trained in comprehensive data, can take advantage of the full feature space effectively to produce highly reliable results. However, when the mask is applied, the performance appears to degrade more noticeably. The scatter plots show a wider spread and a greater deviation from the diagonal, indicating a loss of predictive accuracy.

In contrast, the GNN demonstrates a certain robustness across all tested scenarios, including those with significant masking. Although the performance of the GNN might not be better than that of the FCN, it still consistently provides reasonable results even when substantial portions of the data are masked. This inherent robustness of GNNs is probably due to their ability to model relationships and dependences within the data. This structural understanding allows GNNs to make more informed predictions even when parts of the input are missing or obscured.

The observed differences in performance and robustness between the two networks are fundamentally rooted in their intrinsic architectural designs and how they process information. The FCN offers a significant advantage in terms of computation time. In general, our findings suggest that the choice of architecture should be guided by the specific requirements of the task, balancing accuracy, stability, and computational efficiency.

An interesting direction for future work could be to apply likelihood-free inference methods~\cite{Cranmer_2020,dalmasso2024likelihoodfreefrequentistinferencebridging}, which are particularly well suited to cases where explicit likelihoods are intractable or the analysis is very complicated and computationally expensive.  
Traditional MCMC-based methods can be prohibitively slow despite their ability to naturally provide credible intervals. By adopting likelihood-free inference approaches based on posterior estimation, it becomes possible to generate full posterior distributions, thereby recovering the formal credible intervals as in standard Bayesian analyses, while keeping the computational efficiency of NNs. In the context of GW astronomy, such approaches have been employed to infer source properties of the compact binaries
~\cite{Green:2020hst,Dax:2021tsq,Dax:2022pxd,Dax:2024mcn}, and extended to the SGWB~\cite{Alvey:2023npw,Dimitriou_2024}, enabling a direct comparison with existing MCMC-based results in the literature.

NNs will become essential tools for analyzing the large volumes of upcoming astrometric data with accounting for systematic uncertainties. This paper provides the first step toward the new application of NNs to GW astrometry. While there are still challenges to be addressed, the promising results from this work highlight the potential of NNs in advancing GW astrometry.

% ---------------------------------------------------------
\section*{Acknowledgments}
The authors would like to thank S. Ferraiuolo for discussions at an early stage of the project, and T. S. Yamamoto and H. Takahashi for their helpful discussions. They also acknowledge support from the Research Project No. PID2021-123012NB-C43 and the Spanish Research Agency (Agencia Estatal de Investigaci\'on) through the Grant IFT Centro de Excelencia Severo Ochoa No. CEX2020-001007-S, funded by MCIN/AEI/10.13039/501100011033. The authors also acknowledge the use of the IFT Hydra cluster. M.C. acknowledges support from the Ramón Areces Foundation through the Programa de Ayudas Fundación Ramón Areces para la realización de Tesis Doctorales en Ciencias de la Vida y de la Materia 2023. G.M. acknowledges support from the Ministerio de Universidades through Grant No. FPU20/02857. S.J. acknowledges support from the Agence Nationale de la Recherche (ANR) under Contract No. ANR-22-CE31-0001-01. S.K. is supported by the Spanish Atracci\'on de Talento Contract No. 2023-5A-TIC-28945 granted by Comunidad de Madrid, the I+D Grant No. PID2023-149018NB-C42 funded by MCIN/AEI/10.13039/501100011033, and the Japan Society for the Promotion of Science (JSPS) KAKENHI Grant No. JP20H05853, No. JP23H00110, and No. JP24K00624.

\bibliography{astrometry}

\begin{thebibliography}{58}%
\makeatletter
\providecommand \@ifxundefined [1]{%
 \@ifx{#1\undefined}
}%
\providecommand \@ifnum [1]{%
 \ifnum #1\expandafter \@firstoftwo
 \else \expandafter \@secondoftwo
 \fi
}%
\providecommand \@ifx [1]{%
 \ifx #1\expandafter \@firstoftwo
 \else \expandafter \@secondoftwo
 \fi
}%
\providecommand \natexlab [1]{#1}%
\providecommand \enquote  [1]{``#1''}%
\providecommand \bibnamefont  [1]{#1}%
\providecommand \bibfnamefont [1]{#1}%
\providecommand \citenamefont [1]{#1}%
\providecommand \href@noop [0]{\@secondoftwo}%
\providecommand \href [0]{\begingroup \@sanitize@url \@href}%
\providecommand \@href[1]{\@@startlink{#1}\@@href}%
\providecommand \@@href[1]{\endgroup#1\@@endlink}%
\providecommand \@sanitize@url [0]{\catcode `\\12\catcode `\$12\catcode `\&12\catcode `\#12\catcode `\^12\catcode `\_12\catcode `\%12\relax}%
\providecommand \@@startlink[1]{}%
\providecommand \@@endlink[0]{}%
\providecommand \url  [0]{\begingroup\@sanitize@url \@url }%
\providecommand \@url [1]{\endgroup\@href {#1}{\urlprefix }}%
\providecommand \urlprefix  [0]{URL }%
\providecommand \Eprint [0]{\href }%
\providecommand \doibase [0]{https://doi.org/}%
\providecommand \selectlanguage [0]{\@gobble}%
\providecommand \bibinfo  [0]{\@secondoftwo}%
\providecommand \bibfield  [0]{\@secondoftwo}%
\providecommand \translation [1]{[#1]}%
\providecommand \BibitemOpen [0]{}%
\providecommand \bibitemStop [0]{}%
\providecommand \bibitemNoStop [0]{.\EOS\space}%
\providecommand \EOS [0]{\spacefactor3000\relax}%
\providecommand \BibitemShut  [1]{\csname bibitem#1\endcsname}%
\let\auto@bib@innerbib\@empty
%</preamble>
\bibitem [{\citenamefont {Starobinsky}(1979)}]{Starobinsky:1979ty}%
  \BibitemOpen
  \bibfield  {author} {\bibinfo {author} {\bibfnamefont {A.~A.}\ \bibnamefont {Starobinsky}},\ }\href@noop {} {\bibfield  {journal} {\bibinfo  {journal} {JETP Lett.}\ }\textbf {\bibinfo {volume} {30}},\ \bibinfo {pages} {682} (\bibinfo {year} {1979})}\BibitemShut {NoStop}%
\bibitem [{\citenamefont {Damour}\ and\ \citenamefont {Vilenkin}(2000)}]{Damour:2000wa}%
  \BibitemOpen
  \bibfield  {author} {\bibinfo {author} {\bibfnamefont {T.}~\bibnamefont {Damour}}\ and\ \bibinfo {author} {\bibfnamefont {A.}~\bibnamefont {Vilenkin}},\ }\href {https://doi.org/10.1103/PhysRevLett.85.3761} {\bibfield  {journal} {\bibinfo  {journal} {Phys. Rev. Lett.}\ }\textbf {\bibinfo {volume} {85}},\ \bibinfo {pages} {3761} (\bibinfo {year} {2000})},\ \Eprint {https://arxiv.org/abs/gr-qc/0004075} {arXiv:gr-qc/0004075} \BibitemShut {NoStop}%
\bibitem [{\citenamefont {Kosowsky}\ \emph {et~al.}(1992)\citenamefont {Kosowsky}, \citenamefont {Turner},\ and\ \citenamefont {Watkins}}]{Kosowsky:1991ua}%
  \BibitemOpen
  \bibfield  {author} {\bibinfo {author} {\bibfnamefont {A.}~\bibnamefont {Kosowsky}}, \bibinfo {author} {\bibfnamefont {M.~S.}\ \bibnamefont {Turner}},\ and\ \bibinfo {author} {\bibfnamefont {R.}~\bibnamefont {Watkins}},\ }\href {https://doi.org/10.1103/PhysRevD.45.4514} {\bibfield  {journal} {\bibinfo  {journal} {Phys. Rev. D}\ }\textbf {\bibinfo {volume} {45}},\ \bibinfo {pages} {4514} (\bibinfo {year} {1992})}\BibitemShut {NoStop}%
\bibitem [{\citenamefont {Regimbau}(2011)}]{Regimbau:2011rp}%
  \BibitemOpen
  \bibfield  {author} {\bibinfo {author} {\bibfnamefont {T.}~\bibnamefont {Regimbau}},\ }\href {https://doi.org/10.1088/1674-4527/11/4/001} {\bibfield  {journal} {\bibinfo  {journal} {Res. Astron. Astrophys.}\ }\textbf {\bibinfo {volume} {11}},\ \bibinfo {pages} {369} (\bibinfo {year} {2011})},\ \Eprint {https://arxiv.org/abs/1101.2762} {arXiv:1101.2762 [astro-ph.CO]} \BibitemShut {NoStop}%
\bibitem [{\citenamefont {Caprini}\ and\ \citenamefont {Figueroa}(2018)}]{Caprini:2018mtu}%
  \BibitemOpen
  \bibfield  {author} {\bibinfo {author} {\bibfnamefont {C.}~\bibnamefont {Caprini}}\ and\ \bibinfo {author} {\bibfnamefont {D.~G.}\ \bibnamefont {Figueroa}},\ }\href {https://doi.org/10.1088/1361-6382/aac608} {\bibfield  {journal} {\bibinfo  {journal} {Class. Quant. Grav.}\ }\textbf {\bibinfo {volume} {35}},\ \bibinfo {pages} {163001} (\bibinfo {year} {2018})},\ \Eprint {https://arxiv.org/abs/1801.04268} {arXiv:1801.04268 [astro-ph.CO]} \BibitemShut {NoStop}%
\bibitem [{\citenamefont {Kuroyanagi}\ \emph {et~al.}(2018)\citenamefont {Kuroyanagi}, \citenamefont {Chiba},\ and\ \citenamefont {Takahashi}}]{Kuroyanagi:2018csn}%
  \BibitemOpen
  \bibfield  {author} {\bibinfo {author} {\bibfnamefont {S.}~\bibnamefont {Kuroyanagi}}, \bibinfo {author} {\bibfnamefont {T.}~\bibnamefont {Chiba}},\ and\ \bibinfo {author} {\bibfnamefont {T.}~\bibnamefont {Takahashi}},\ }\href {https://doi.org/10.1088/1475-7516/2018/11/038} {\bibfield  {journal} {\bibinfo  {journal} {JCAP}\ }\textbf {\bibinfo {volume} {11}},\ \bibinfo {pages} {038}},\ \Eprint {https://arxiv.org/abs/1807.00786} {arXiv:1807.00786 [astro-ph.CO]} \BibitemShut {NoStop}%
\bibitem [{\citenamefont {Christensen}(2018)}]{Christensen_2018}%
  \BibitemOpen
  \bibfield  {author} {\bibinfo {author} {\bibfnamefont {N.}~\bibnamefont {Christensen}},\ }\href {https://doi.org/10.1088/1361-6633/aae6b5} {\bibfield  {journal} {\bibinfo  {journal} {Reports on Progress in Physics}\ }\textbf {\bibinfo {volume} {82}},\ \bibinfo {pages} {016903} (\bibinfo {year} {2018})}\BibitemShut {NoStop}%
\bibitem [{\citenamefont {Colpi}\ \emph {et~al.}(2024)\citenamefont {Colpi} \emph {et~al.}}]{Colpi:2024xhw}%
  \BibitemOpen
  \bibfield  {author} {\bibinfo {author} {\bibfnamefont {M.}~\bibnamefont {Colpi}} \emph {et~al.},\ }\href@noop {} {\  (\bibinfo {year} {2024})},\ \Eprint {https://arxiv.org/abs/2402.07571} {arXiv:2402.07571 [astro-ph.CO]} \BibitemShut {NoStop}%
\bibitem [{\citenamefont {Kawamura}\ \emph {et~al.}(2021)\citenamefont {Kawamura} \emph {et~al.}}]{Kawamura:2020pcg}%
  \BibitemOpen
  \bibfield  {author} {\bibinfo {author} {\bibfnamefont {S.}~\bibnamefont {Kawamura}} \emph {et~al.},\ }\href {https://doi.org/10.1093/ptep/ptab019} {\bibfield  {journal} {\bibinfo  {journal} {PTEP}\ }\textbf {\bibinfo {volume} {2021}},\ \bibinfo {pages} {05A105} (\bibinfo {year} {2021})},\ \Eprint {https://arxiv.org/abs/2006.13545} {arXiv:2006.13545 [gr-qc]} \BibitemShut {NoStop}%
\bibitem [{\citenamefont {Abbott}\ \emph {et~al.}(2017)\citenamefont {Abbott}, \citenamefont {Abbott}, \citenamefont {Abbott}, \citenamefont {Abernathy}, \citenamefont {Ackley}, \citenamefont {Adams}, \citenamefont {Addesso}, \citenamefont {Adhikari},\ and\ \citenamefont {at~al}}]{Abbott_2017}%
  \BibitemOpen
  \bibfield  {author} {\bibinfo {author} {\bibfnamefont {B.~P.}\ \bibnamefont {Abbott}}, \bibinfo {author} {\bibfnamefont {R.}~\bibnamefont {Abbott}}, \bibinfo {author} {\bibfnamefont {T.~D.}\ \bibnamefont {Abbott}}, \bibinfo {author} {\bibfnamefont {M.~R.}\ \bibnamefont {Abernathy}}, \bibinfo {author} {\bibfnamefont {K.}~\bibnamefont {Ackley}}, \bibinfo {author} {\bibfnamefont {C.}~\bibnamefont {Adams}}, \bibinfo {author} {\bibfnamefont {P.}~\bibnamefont {Addesso}}, \bibinfo {author} {\bibfnamefont {R.~X.}\ \bibnamefont {Adhikari}},\ and\ \bibinfo {author} {\bibnamefont {at~al}},\ }\href {https://doi.org/10.1088/1361-6382/aa51f4} {\bibfield  {journal} {\bibinfo  {journal} {Classical and Quantum Gravity}\ }\textbf {\bibinfo {volume} {34}},\ \bibinfo {pages} {044001} (\bibinfo {year} {2017})}\BibitemShut {NoStop}%
\bibitem [{\citenamefont {Domcke}(2023)}]{Domcke:2023qle}%
  \BibitemOpen
  \bibfield  {author} {\bibinfo {author} {\bibfnamefont {V.}~\bibnamefont {Domcke}},\ }in\ \href@noop {} {\emph {\bibinfo {booktitle} {{57th Rencontres de Moriond on Electroweak Interactions and Unified Theories}}}}\ (\bibinfo {year} {2023})\ \Eprint {https://arxiv.org/abs/2306.04496} {arXiv:2306.04496 [gr-qc]} \BibitemShut {NoStop}%
\bibitem [{\citenamefont {Hazumi}\ \emph {et~al.}(2020)\citenamefont {Hazumi} \emph {et~al.}}]{LiteBIRD:2020khw}%
  \BibitemOpen
  \bibfield  {author} {\bibinfo {author} {\bibfnamefont {M.}~\bibnamefont {Hazumi}} \emph {et~al.} (\bibinfo {collaboration} {LiteBIRD}),\ }\href {https://doi.org/10.1117/12.2563050} {\bibfield  {journal} {\bibinfo  {journal} {Proc. SPIE Int. Soc. Opt. Eng.}\ }\textbf {\bibinfo {volume} {11443}},\ \bibinfo {pages} {114432F} (\bibinfo {year} {2020})},\ \Eprint {https://arxiv.org/abs/2101.12449} {arXiv:2101.12449 [astro-ph.IM]} \BibitemShut {NoStop}%
\bibitem [{\citenamefont {Agazie}\ \emph {et~al.}(2024)\citenamefont {Agazie} \emph {et~al.}}]{InternationalPulsarTimingArray:2023mzf}%
  \BibitemOpen
  \bibfield  {author} {\bibinfo {author} {\bibfnamefont {G.}~\bibnamefont {Agazie}} \emph {et~al.} (\bibinfo {collaboration} {International Pulsar Timing Array}),\ }\href {https://doi.org/10.3847/1538-4357/ad36be} {\bibfield  {journal} {\bibinfo  {journal} {Astrophys. J.}\ }\textbf {\bibinfo {volume} {966}},\ \bibinfo {pages} {105} (\bibinfo {year} {2024})},\ \Eprint {https://arxiv.org/abs/2309.00693} {arXiv:2309.00693 [astro-ph.HE]} \BibitemShut {NoStop}%
\bibitem [{\citenamefont {Pyne}\ \emph {et~al.}(1996)\citenamefont {Pyne}, \citenamefont {Gwinn}, \citenamefont {Birkinshaw}, \citenamefont {Eubanks},\ and\ \citenamefont {Matsakis}}]{Pyne:1995iy}%
  \BibitemOpen
  \bibfield  {author} {\bibinfo {author} {\bibfnamefont {T.}~\bibnamefont {Pyne}}, \bibinfo {author} {\bibfnamefont {C.~R.}\ \bibnamefont {Gwinn}}, \bibinfo {author} {\bibfnamefont {M.}~\bibnamefont {Birkinshaw}}, \bibinfo {author} {\bibfnamefont {T.~M.}\ \bibnamefont {Eubanks}},\ and\ \bibinfo {author} {\bibfnamefont {D.~N.}\ \bibnamefont {Matsakis}},\ }\href {https://doi.org/10.1086/177443} {\bibfield  {journal} {\bibinfo  {journal} {Astrophys. J.}\ }\textbf {\bibinfo {volume} {465}},\ \bibinfo {pages} {566} (\bibinfo {year} {1996})},\ \Eprint {https://arxiv.org/abs/astro-ph/9507030} {arXiv:astro-ph/9507030} \BibitemShut {NoStop}%
\bibitem [{\citenamefont {Jaffe}(2004)}]{Jaffe:2004it}%
  \BibitemOpen
  \bibfield  {author} {\bibinfo {author} {\bibfnamefont {A.~H.}\ \bibnamefont {Jaffe}},\ }\href {https://doi.org/10.1016/j.newar.2004.09.018} {\bibfield  {journal} {\bibinfo  {journal} {New Astron. Rev.}\ }\textbf {\bibinfo {volume} {48}},\ \bibinfo {pages} {1483} (\bibinfo {year} {2004})},\ \Eprint {https://arxiv.org/abs/astro-ph/0409637} {arXiv:astro-ph/0409637} \BibitemShut {NoStop}%
\bibitem [{\citenamefont {Book}\ and\ \citenamefont {Flanagan}(2011)}]{PhysRevD.83.024024}%
  \BibitemOpen
  \bibfield  {author} {\bibinfo {author} {\bibfnamefont {L.~G.}\ \bibnamefont {Book}}\ and\ \bibinfo {author} {\bibfnamefont {E.~E.}\ \bibnamefont {Flanagan}},\ }\href {https://doi.org/10.1103/PhysRevD.83.024024} {\bibfield  {journal} {\bibinfo  {journal} {Phys. Rev. D}\ }\textbf {\bibinfo {volume} {83}},\ \bibinfo {pages} {024024} (\bibinfo {year} {2011})}\BibitemShut {NoStop}%
\bibitem [{\citenamefont {Mihaylov}\ \emph {et~al.}(2018)\citenamefont {Mihaylov}, \citenamefont {Moore}, \citenamefont {Gair}, \citenamefont {Lasenby},\ and\ \citenamefont {Gilmore}}]{Mihaylov:2018uqm}%
  \BibitemOpen
  \bibfield  {author} {\bibinfo {author} {\bibfnamefont {D.~P.}\ \bibnamefont {Mihaylov}}, \bibinfo {author} {\bibfnamefont {C.~J.}\ \bibnamefont {Moore}}, \bibinfo {author} {\bibfnamefont {J.~R.}\ \bibnamefont {Gair}}, \bibinfo {author} {\bibfnamefont {A.}~\bibnamefont {Lasenby}},\ and\ \bibinfo {author} {\bibfnamefont {G.}~\bibnamefont {Gilmore}},\ }\href {https://doi.org/10.1103/PhysRevD.97.124058} {\bibfield  {journal} {\bibinfo  {journal} {Phys. Rev. D}\ }\textbf {\bibinfo {volume} {97}},\ \bibinfo {pages} {124058} (\bibinfo {year} {2018})},\ \Eprint {https://arxiv.org/abs/1804.00660} {arXiv:1804.00660 [gr-qc]} \BibitemShut {NoStop}%
\bibitem [{\citenamefont {Mihaylov}\ \emph {et~al.}(2020)\citenamefont {Mihaylov}, \citenamefont {Moore}, \citenamefont {Gair}, \citenamefont {Lasenby},\ and\ \citenamefont {Gilmore}}]{Mihaylov:2019lft}%
  \BibitemOpen
  \bibfield  {author} {\bibinfo {author} {\bibfnamefont {D.~P.}\ \bibnamefont {Mihaylov}}, \bibinfo {author} {\bibfnamefont {C.~J.}\ \bibnamefont {Moore}}, \bibinfo {author} {\bibfnamefont {J.}~\bibnamefont {Gair}}, \bibinfo {author} {\bibfnamefont {A.}~\bibnamefont {Lasenby}},\ and\ \bibinfo {author} {\bibfnamefont {G.}~\bibnamefont {Gilmore}},\ }\href {https://doi.org/10.1103/PhysRevD.101.024038} {\bibfield  {journal} {\bibinfo  {journal} {Phys. Rev. D}\ }\textbf {\bibinfo {volume} {101}},\ \bibinfo {pages} {024038} (\bibinfo {year} {2020})},\ \Eprint {https://arxiv.org/abs/1911.10356} {arXiv:1911.10356 [gr-qc]} \BibitemShut {NoStop}%
\bibitem [{\citenamefont {Prusti}\ \emph {et~al.}(2016)\citenamefont {Prusti} \emph {et~al.}}]{Gaia:2016zol}%
  \BibitemOpen
  \bibfield  {author} {\bibinfo {author} {\bibfnamefont {T.}~\bibnamefont {Prusti}} \emph {et~al.} (\bibinfo {collaboration} {Gaia}),\ }\href {https://doi.org/10.1051/0004-6361/201629272} {\bibfield  {journal} {\bibinfo  {journal} {Astron. Astrophys.}\ }\textbf {\bibinfo {volume} {595}},\ \bibinfo {pages} {A1} (\bibinfo {year} {2016})},\ \Eprint {https://arxiv.org/abs/1609.04153} {arXiv:1609.04153 [astro-ph.IM]} \BibitemShut {NoStop}%
\bibitem [{\citenamefont {Gwinn}\ \emph {et~al.}(1997)\citenamefont {Gwinn}, \citenamefont {Eubanks}, \citenamefont {Pyne}, \citenamefont {Birkinshaw},\ and\ \citenamefont {Matsakis}}]{Gwinn:1996gv}%
  \BibitemOpen
  \bibfield  {author} {\bibinfo {author} {\bibfnamefont {C.~R.}\ \bibnamefont {Gwinn}}, \bibinfo {author} {\bibfnamefont {T.~M.}\ \bibnamefont {Eubanks}}, \bibinfo {author} {\bibfnamefont {T.}~\bibnamefont {Pyne}}, \bibinfo {author} {\bibfnamefont {M.}~\bibnamefont {Birkinshaw}},\ and\ \bibinfo {author} {\bibfnamefont {D.~N.}\ \bibnamefont {Matsakis}},\ }\href {https://doi.org/10.1086/304424} {\bibfield  {journal} {\bibinfo  {journal} {Astrophys. J.}\ }\textbf {\bibinfo {volume} {485}},\ \bibinfo {pages} {87} (\bibinfo {year} {1997})},\ \Eprint {https://arxiv.org/abs/astro-ph/9610086} {arXiv:astro-ph/9610086} \BibitemShut {NoStop}%
\bibitem [{\citenamefont {Titov}\ \emph {et~al.}(2011)\citenamefont {Titov}, \citenamefont {Lambert},\ and\ \citenamefont {Gontier}}]{Titov:2010zn}%
  \BibitemOpen
  \bibfield  {author} {\bibinfo {author} {\bibfnamefont {O.}~\bibnamefont {Titov}}, \bibinfo {author} {\bibfnamefont {S.~B.}\ \bibnamefont {Lambert}},\ and\ \bibinfo {author} {\bibfnamefont {A.~M.}\ \bibnamefont {Gontier}},\ }\href {https://doi.org/10.1051/0004-6361/201015718} {\bibfield  {journal} {\bibinfo  {journal} {Astron. Astrophys.}\ }\textbf {\bibinfo {volume} {529}},\ \bibinfo {pages} {A91} (\bibinfo {year} {2011})},\ \Eprint {https://arxiv.org/abs/1009.3698} {arXiv:1009.3698 [astro-ph.CO]} \BibitemShut {NoStop}%
\bibitem [{\citenamefont {Darling}\ \emph {et~al.}(2018)\citenamefont {Darling}, \citenamefont {Truebenbach},\ and\ \citenamefont {Paine}}]{Darling_2018}%
  \BibitemOpen
  \bibfield  {author} {\bibinfo {author} {\bibfnamefont {J.}~\bibnamefont {Darling}}, \bibinfo {author} {\bibfnamefont {A.~E.}\ \bibnamefont {Truebenbach}},\ and\ \bibinfo {author} {\bibfnamefont {J.}~\bibnamefont {Paine}},\ }\href {https://doi.org/10.3847/1538-4357/aac772} {\bibfield  {journal} {\bibinfo  {journal} {The Astrophysical Journal}\ }\textbf {\bibinfo {volume} {861}},\ \bibinfo {pages} {113} (\bibinfo {year} {2018})}\BibitemShut {NoStop}%
\bibitem [{\citenamefont {Jaraba}\ \emph {et~al.}(2023)\citenamefont {Jaraba}, \citenamefont {Garc\'\i{}a-Bellido}, \citenamefont {Kuroyanagi}, \citenamefont {Ferraiuolo},\ and\ \citenamefont {Braglia}}]{Jaraba:2023djs}%
  \BibitemOpen
  \bibfield  {author} {\bibinfo {author} {\bibfnamefont {S.}~\bibnamefont {Jaraba}}, \bibinfo {author} {\bibfnamefont {J.}~\bibnamefont {Garc\'\i{}a-Bellido}}, \bibinfo {author} {\bibfnamefont {S.}~\bibnamefont {Kuroyanagi}}, \bibinfo {author} {\bibfnamefont {S.}~\bibnamefont {Ferraiuolo}},\ and\ \bibinfo {author} {\bibfnamefont {M.}~\bibnamefont {Braglia}},\ }\href {https://doi.org/10.1093/mnras/stad2141} {\bibfield  {journal} {\bibinfo  {journal} {Mon. Not. Roy. Astron. Soc.}\ }\textbf {\bibinfo {volume} {524}},\ \bibinfo {pages} {3609} (\bibinfo {year} {2023})},\ \Eprint {https://arxiv.org/abs/2304.06350} {arXiv:2304.06350 [astro-ph.CO]} \BibitemShut {NoStop}%
\bibitem [{\citenamefont {Darling}(2025)}]{Darling:2024myz}%
  \BibitemOpen
  \bibfield  {author} {\bibinfo {author} {\bibfnamefont {J.}~\bibnamefont {Darling}},\ }\href {https://doi.org/10.3847/2041-8213/adbf0d} {\bibfield  {journal} {\bibinfo  {journal} {Astrophys. J. Lett.}\ }\textbf {\bibinfo {volume} {982}},\ \bibinfo {pages} {L46} (\bibinfo {year} {2025})},\ \Eprint {https://arxiv.org/abs/2412.08605} {arXiv:2412.08605 [astro-ph.CO]} \BibitemShut {NoStop}%
\bibitem [{\citenamefont {Yeh}\ \emph {et~al.}(2022)\citenamefont {Yeh}, \citenamefont {Shelton}, \citenamefont {Olive},\ and\ \citenamefont {Fields}}]{Yeh:2022heq}%
  \BibitemOpen
  \bibfield  {author} {\bibinfo {author} {\bibfnamefont {T.-H.}\ \bibnamefont {Yeh}}, \bibinfo {author} {\bibfnamefont {J.}~\bibnamefont {Shelton}}, \bibinfo {author} {\bibfnamefont {K.~A.}\ \bibnamefont {Olive}},\ and\ \bibinfo {author} {\bibfnamefont {B.~D.}\ \bibnamefont {Fields}},\ }\href {https://doi.org/10.1088/1475-7516/2022/10/046} {\bibfield  {journal} {\bibinfo  {journal} {Journal of Cosmology and Astroparticle Physics}\ }\textbf {\bibinfo {volume} {10}},\ \bibinfo {pages} {046}},\ \Eprint {https://arxiv.org/abs/2207.13133} {arXiv:2207.13133 [astro-ph.CO]} \BibitemShut {NoStop}%
\bibitem [{\citenamefont {Kite}\ \emph {et~al.}(2021)\citenamefont {Kite}, \citenamefont {Ravenni}, \citenamefont {Patil},\ and\ \citenamefont {Chluba}}]{Kite:2020uix}%
  \BibitemOpen
  \bibfield  {author} {\bibinfo {author} {\bibfnamefont {T.}~\bibnamefont {Kite}}, \bibinfo {author} {\bibfnamefont {A.}~\bibnamefont {Ravenni}}, \bibinfo {author} {\bibfnamefont {S.~P.}\ \bibnamefont {Patil}},\ and\ \bibinfo {author} {\bibfnamefont {J.}~\bibnamefont {Chluba}},\ }\href {https://doi.org/10.1093/mnras/stab1558} {\bibfield  {journal} {\bibinfo  {journal} {Mon. Not. Roy. Astron. Soc.}\ }\textbf {\bibinfo {volume} {505}},\ \bibinfo {pages} {4396} (\bibinfo {year} {2021})},\ \Eprint {https://arxiv.org/abs/2010.00040} {arXiv:2010.00040 [astro-ph.CO]} \BibitemShut {NoStop}%
\bibitem [{\citenamefont {Wang}\ \emph {et~al.}(2022)\citenamefont {Wang}, \citenamefont {Pardo}, \citenamefont {Chang},\ and\ \citenamefont {Dor\'e}}]{Wang:2022sxn}%
  \BibitemOpen
  \bibfield  {author} {\bibinfo {author} {\bibfnamefont {Y.}~\bibnamefont {Wang}}, \bibinfo {author} {\bibfnamefont {K.}~\bibnamefont {Pardo}}, \bibinfo {author} {\bibfnamefont {T.-C.}\ \bibnamefont {Chang}},\ and\ \bibinfo {author} {\bibfnamefont {O.}~\bibnamefont {Dor\'e}},\ }\href {https://doi.org/10.1103/PhysRevD.106.084006} {\bibfield  {journal} {\bibinfo  {journal} {Phys. Rev. D}\ }\textbf {\bibinfo {volume} {106}},\ \bibinfo {pages} {084006} (\bibinfo {year} {2022})},\ \Eprint {https://arxiv.org/abs/2205.07962} {arXiv:2205.07962 [gr-qc]} \BibitemShut {NoStop}%
\bibitem [{\citenamefont {Pardo}\ \emph {et~al.}(2023)\citenamefont {Pardo}, \citenamefont {Chang}, \citenamefont {Dor\'e},\ and\ \citenamefont {Wang}}]{Pardo:2023cag}%
  \BibitemOpen
  \bibfield  {author} {\bibinfo {author} {\bibfnamefont {K.}~\bibnamefont {Pardo}}, \bibinfo {author} {\bibfnamefont {T.-C.}\ \bibnamefont {Chang}}, \bibinfo {author} {\bibfnamefont {O.}~\bibnamefont {Dor\'e}},\ and\ \bibinfo {author} {\bibfnamefont {Y.}~\bibnamefont {Wang}},\ }\href@noop {} {\  (\bibinfo {year} {2023})},\ \Eprint {https://arxiv.org/abs/2306.14968} {arXiv:2306.14968 [astro-ph.GA]} \BibitemShut {NoStop}%
\bibitem [{\citenamefont {Malbet}\ \emph {et~al.}(2022)\citenamefont {Malbet} \emph {et~al.}}]{Malbet:2022lll}%
  \BibitemOpen
  \bibfield  {author} {\bibinfo {author} {\bibfnamefont {F.}~\bibnamefont {Malbet}} \emph {et~al.},\ }in\ \href@noop {} {\emph {\bibinfo {booktitle} {{SPIE Astronomical Telescopes + Instrumentation 2022}}}}\ (\bibinfo {year} {2022})\ \Eprint {https://arxiv.org/abs/2207.12540} {arXiv:2207.12540 [astro-ph.IM]} \BibitemShut {NoStop}%
\bibitem [{\citenamefont {García-Bellido}\ \emph {et~al.}(2021)\citenamefont {García-Bellido}, \citenamefont {Murayama},\ and\ \citenamefont {White}}]{Garcia-Bellido:2021zgu}%
  \BibitemOpen
  \bibfield  {author} {\bibinfo {author} {\bibfnamefont {J.}~\bibnamefont {García-Bellido}}, \bibinfo {author} {\bibfnamefont {H.}~\bibnamefont {Murayama}},\ and\ \bibinfo {author} {\bibfnamefont {G.}~\bibnamefont {White}},\ }\href {https://doi.org/10.1088/1475-7516/2021/12/023} {\bibfield  {journal} {\bibinfo  {journal} {JCAP}\ }\textbf {\bibinfo {volume} {12}}\bibfield  {number} {\bibinfo  {number} { (12)},\ \bibinfo {pages} {023}},\ }\Eprint {https://arxiv.org/abs/2104.04778} {arXiv:2104.04778 [hep-ph]} \BibitemShut {NoStop}%
\bibitem [{\citenamefont {Moore}\ \emph {et~al.}(2017)\citenamefont {Moore}, \citenamefont {Mihaylov}, \citenamefont {Lasenby},\ and\ \citenamefont {Gilmore}}]{Moore:2017ity}%
  \BibitemOpen
  \bibfield  {author} {\bibinfo {author} {\bibfnamefont {C.~J.}\ \bibnamefont {Moore}}, \bibinfo {author} {\bibfnamefont {D.~P.}\ \bibnamefont {Mihaylov}}, \bibinfo {author} {\bibfnamefont {A.}~\bibnamefont {Lasenby}},\ and\ \bibinfo {author} {\bibfnamefont {G.}~\bibnamefont {Gilmore}},\ }\href {https://doi.org/10.1103/PhysRevLett.119.261102} {\bibfield  {journal} {\bibinfo  {journal} {Phys. Rev. Lett.}\ }\textbf {\bibinfo {volume} {119}},\ \bibinfo {pages} {261102} (\bibinfo {year} {2017})},\ \Eprint {https://arxiv.org/abs/1707.06239} {arXiv:1707.06239 [astro-ph.IM]} \BibitemShut {NoStop}%
\bibitem [{\citenamefont {Qin}\ \emph {et~al.}(2019)\citenamefont {Qin}, \citenamefont {Boddy}, \citenamefont {Kamionkowski},\ and\ \citenamefont {Dai}}]{Qin:2018yhy}%
  \BibitemOpen
  \bibfield  {author} {\bibinfo {author} {\bibfnamefont {W.}~\bibnamefont {Qin}}, \bibinfo {author} {\bibfnamefont {K.~K.}\ \bibnamefont {Boddy}}, \bibinfo {author} {\bibfnamefont {M.}~\bibnamefont {Kamionkowski}},\ and\ \bibinfo {author} {\bibfnamefont {L.}~\bibnamefont {Dai}},\ }\href {https://doi.org/10.1103/PhysRevD.99.063002} {\bibfield  {journal} {\bibinfo  {journal} {Phys. Rev. D}\ }\textbf {\bibinfo {volume} {99}},\ \bibinfo {pages} {063002} (\bibinfo {year} {2019})},\ \Eprint {https://arxiv.org/abs/1810.02369} {arXiv:1810.02369 [astro-ph.CO]} \BibitemShut {NoStop}%
\bibitem [{\citenamefont {{\c{C}}al{\i}{\c{s}}kan}\ \emph {et~al.}(2024)\citenamefont {{\c{C}}al{\i}{\c{s}}kan}, \citenamefont {Chen}, \citenamefont {Dai}, \citenamefont {Anil~Kumar}, \citenamefont {Stomberg},\ and\ \citenamefont {Xue}}]{Caliskan:2023cqm}%
  \BibitemOpen
  \bibfield  {author} {\bibinfo {author} {\bibfnamefont {M.}~\bibnamefont {{\c{C}}al{\i}{\c{s}}kan}}, \bibinfo {author} {\bibfnamefont {Y.}~\bibnamefont {Chen}}, \bibinfo {author} {\bibfnamefont {L.}~\bibnamefont {Dai}}, \bibinfo {author} {\bibfnamefont {N.}~\bibnamefont {Anil~Kumar}}, \bibinfo {author} {\bibfnamefont {I.}~\bibnamefont {Stomberg}},\ and\ \bibinfo {author} {\bibfnamefont {X.}~\bibnamefont {Xue}},\ }\href {https://doi.org/10.1088/1475-7516/2024/05/030} {\bibfield  {journal} {\bibinfo  {journal} {JCAP}\ }\textbf {\bibinfo {volume} {05}},\ \bibinfo {pages} {030}},\ \Eprint {https://arxiv.org/abs/2312.03069} {arXiv:2312.03069 [gr-qc]} \BibitemShut {NoStop}%
\bibitem [{\citenamefont {Inomata}\ \emph {et~al.}(2024)\citenamefont {Inomata}, \citenamefont {Kamionkowski}, \citenamefont {Toral},\ and\ \citenamefont {Taylor}}]{Inomata:2024kzr}%
  \BibitemOpen
  \bibfield  {author} {\bibinfo {author} {\bibfnamefont {K.}~\bibnamefont {Inomata}}, \bibinfo {author} {\bibfnamefont {M.}~\bibnamefont {Kamionkowski}}, \bibinfo {author} {\bibfnamefont {C.~M.}\ \bibnamefont {Toral}},\ and\ \bibinfo {author} {\bibfnamefont {S.~R.}\ \bibnamefont {Taylor}},\ }\href {https://doi.org/10.1103/PhysRevD.110.063547} {\bibfield  {journal} {\bibinfo  {journal} {Phys. Rev. D}\ }\textbf {\bibinfo {volume} {110}},\ \bibinfo {pages} {063547} (\bibinfo {year} {2024})},\ \Eprint {https://arxiv.org/abs/2406.00096} {arXiv:2406.00096 [astro-ph.CO]} \BibitemShut {NoStop}%
\bibitem [{\citenamefont {Cruz}\ \emph {et~al.}(2024)\citenamefont {Cruz}, \citenamefont {Malhotra}, \citenamefont {Tasinato},\ and\ \citenamefont {Zavala}}]{Cruz:2024diu}%
  \BibitemOpen
  \bibfield  {author} {\bibinfo {author} {\bibfnamefont {N.~M.~J.}\ \bibnamefont {Cruz}}, \bibinfo {author} {\bibfnamefont {A.}~\bibnamefont {Malhotra}}, \bibinfo {author} {\bibfnamefont {G.}~\bibnamefont {Tasinato}},\ and\ \bibinfo {author} {\bibfnamefont {I.}~\bibnamefont {Zavala}},\ }\href@noop {} {\  (\bibinfo {year} {2024})},\ \Eprint {https://arxiv.org/abs/2412.14010} {arXiv:2412.14010 [astro-ph.CO]} \BibitemShut {NoStop}%
\bibitem [{\citenamefont {Maggiore}(2007)}]{Maggiore:2007ulw}%
  \BibitemOpen
  \bibfield  {author} {\bibinfo {author} {\bibfnamefont {M.}~\bibnamefont {Maggiore}},\ }\href {https://doi.org/10.1093/acprof:oso/9780198570745.001.0001} {\emph {\bibinfo {title} {{Gravitational Waves. Vol. 1: Theory and Experiments}}}}\ (\bibinfo  {publisher} {Oxford University Press},\ \bibinfo {year} {2007})\BibitemShut {NoStop}%
\bibitem [{\citenamefont {Romano}\ and\ \citenamefont {Cornish}(2017)}]{Romano:2016dpx}%
  \BibitemOpen
  \bibfield  {author} {\bibinfo {author} {\bibfnamefont {J.~D.}\ \bibnamefont {Romano}}\ and\ \bibinfo {author} {\bibfnamefont {N.~J.}\ \bibnamefont {Cornish}},\ }\href {https://doi.org/10.1007/s41114-017-0004-1} {\bibfield  {journal} {\bibinfo  {journal} {Living Rev. Rel.}\ }\textbf {\bibinfo {volume} {20}},\ \bibinfo {pages} {2} (\bibinfo {year} {2017})},\ \Eprint {https://arxiv.org/abs/1608.06889} {arXiv:1608.06889 [gr-qc]} \BibitemShut {NoStop}%
\bibitem [{pyg()}]{pygaia}%
  \BibitemOpen
  \href@noop {} {}\bibinfo {howpublished} {\url{https://github.com/agabrown/PyGaia}}\BibitemShut {NoStop}%
\bibitem [{\citenamefont {Storey-Fisher}\ \emph {et~al.}(2024)\citenamefont {Storey-Fisher}, \citenamefont {Hogg}, \citenamefont {Rix}, \citenamefont {Eilers}, \citenamefont {Fabbian}, \citenamefont {Blanton},\ and\ \citenamefont {Alonso}}]{Storey-Fisher:2023gca}%
  \BibitemOpen
  \bibfield  {author} {\bibinfo {author} {\bibfnamefont {K.}~\bibnamefont {Storey-Fisher}}, \bibinfo {author} {\bibfnamefont {D.~W.}\ \bibnamefont {Hogg}}, \bibinfo {author} {\bibfnamefont {H.-W.}\ \bibnamefont {Rix}}, \bibinfo {author} {\bibfnamefont {A.-C.}\ \bibnamefont {Eilers}}, \bibinfo {author} {\bibfnamefont {G.}~\bibnamefont {Fabbian}}, \bibinfo {author} {\bibfnamefont {M.~R.}\ \bibnamefont {Blanton}},\ and\ \bibinfo {author} {\bibfnamefont {D.}~\bibnamefont {Alonso}},\ }\href {https://doi.org/10.3847/1538-4357/ad1328} {\bibfield  {journal} {\bibinfo  {journal} {Astrophys. J.}\ }\textbf {\bibinfo {volume} {964}},\ \bibinfo {pages} {69} (\bibinfo {year} {2024})},\ \Eprint {https://arxiv.org/abs/2306.17749} {arXiv:2306.17749 [astro-ph.GA]} \BibitemShut {NoStop}%
\bibitem [{gai()}]{gaiadoc}%
  \BibitemOpen
  \href@noop {} {}\bibinfo {howpublished} {\url{https://www.cosmos.esa.int/web/gaia/science-performance}}\BibitemShut {NoStop}%
\bibitem [{\citenamefont {Mignard}\ and\ \citenamefont {Klioner}(2012)}]{Mignard:2012xm}%
  \BibitemOpen
  \bibfield  {author} {\bibinfo {author} {\bibfnamefont {F.}~\bibnamefont {Mignard}}\ and\ \bibinfo {author} {\bibfnamefont {S.}~\bibnamefont {Klioner}},\ }\href {https://doi.org/10.1051/0004-6361/201219927} {\bibfield  {journal} {\bibinfo  {journal} {Astron. Astrophys.}\ }\textbf {\bibinfo {volume} {547}},\ \bibinfo {pages} {A59} (\bibinfo {year} {2012})},\ \Eprint {https://arxiv.org/abs/1207.0025} {arXiv:1207.0025 [astro-ph.IM]} \BibitemShut {NoStop}%
\bibitem [{\citenamefont {Zhou}\ \emph {et~al.}(2020)\citenamefont {Zhou}, \citenamefont {Cui}, \citenamefont {Hu}, \citenamefont {Zhang}, \citenamefont {Yang}, \citenamefont {Liu}, \citenamefont {Wang}, \citenamefont {Li},\ and\ \citenamefont {Sun}}]{ZHOU202057}%
  \BibitemOpen
  \bibfield  {author} {\bibinfo {author} {\bibfnamefont {J.}~\bibnamefont {Zhou}}, \bibinfo {author} {\bibfnamefont {G.}~\bibnamefont {Cui}}, \bibinfo {author} {\bibfnamefont {S.}~\bibnamefont {Hu}}, \bibinfo {author} {\bibfnamefont {Z.}~\bibnamefont {Zhang}}, \bibinfo {author} {\bibfnamefont {C.}~\bibnamefont {Yang}}, \bibinfo {author} {\bibfnamefont {Z.}~\bibnamefont {Liu}}, \bibinfo {author} {\bibfnamefont {L.}~\bibnamefont {Wang}}, \bibinfo {author} {\bibfnamefont {C.}~\bibnamefont {Li}},\ and\ \bibinfo {author} {\bibfnamefont {M.}~\bibnamefont {Sun}},\ }\href {https://doi.org/https://doi.org/10.1016/j.aiopen.2021.01.001} {\bibfield  {journal} {\bibinfo  {journal} {AI Open}\ }\textbf {\bibinfo {volume} {1}},\ \bibinfo {pages} {57} (\bibinfo {year} {2020})}\BibitemShut {NoStop}%
\bibitem [{\citenamefont {Sanchez-Lengeling}\ \emph {et~al.}(2021)\citenamefont {Sanchez-Lengeling}, \citenamefont {Reif}, \citenamefont {Pearce},\ and\ \citenamefont {Wiltschko}}]{sanchez-lengeling2021a}%
  \BibitemOpen
  \bibfield  {author} {\bibinfo {author} {\bibfnamefont {B.}~\bibnamefont {Sanchez-Lengeling}}, \bibinfo {author} {\bibfnamefont {E.}~\bibnamefont {Reif}}, \bibinfo {author} {\bibfnamefont {A.}~\bibnamefont {Pearce}},\ and\ \bibinfo {author} {\bibfnamefont {A.~B.}\ \bibnamefont {Wiltschko}},\ }\bibfield  {journal} {\bibinfo  {journal} {Distill}\ }\href {https://doi.org/10.23915/distill.00033} {10.23915/distill.00033} (\bibinfo {year} {2021}),\ \bibinfo {note} {https://distill.pub/2021/gnn-intro}\BibitemShut {NoStop}%
\bibitem [{\citenamefont {Akiba}\ \emph {et~al.}(2019)\citenamefont {Akiba}, \citenamefont {Sano}, \citenamefont {Yanase}, \citenamefont {Ohta},\ and\ \citenamefont {Koyama}}]{optuna_2019}%
  \BibitemOpen
  \bibfield  {author} {\bibinfo {author} {\bibfnamefont {T.}~\bibnamefont {Akiba}}, \bibinfo {author} {\bibfnamefont {S.}~\bibnamefont {Sano}}, \bibinfo {author} {\bibfnamefont {T.}~\bibnamefont {Yanase}}, \bibinfo {author} {\bibfnamefont {T.}~\bibnamefont {Ohta}},\ and\ \bibinfo {author} {\bibfnamefont {M.}~\bibnamefont {Koyama}},\ }in\ \href {https://doi.org/10.1145/3292500.3330701} {\emph {\bibinfo {booktitle} {Proceedings of the 25th ACM SIGKDD International Conference on Knowledge Discovery \& Data Mining}}},\ \bibinfo {series and number} {KDD '19}\ (\bibinfo  {publisher} {Association for Computing Machinery},\ \bibinfo {address} {New York, NY, USA},\ \bibinfo {year} {2019})\ p.\ \bibinfo {pages} {2623–2631}\BibitemShut {NoStop}%
\bibitem [{\citenamefont {Sivakumar}\ \emph {et~al.}(2024)\citenamefont {Sivakumar}, \citenamefont {Parthasarathy},\ and\ \citenamefont {Padmapriya}}]{sivakumar2024trade}%
  \BibitemOpen
  \bibfield  {author} {\bibinfo {author} {\bibfnamefont {M.}~\bibnamefont {Sivakumar}}, \bibinfo {author} {\bibfnamefont {S.}~\bibnamefont {Parthasarathy}},\ and\ \bibinfo {author} {\bibfnamefont {T.}~\bibnamefont {Padmapriya}},\ }\href@noop {} {\bibfield  {journal} {\bibinfo  {journal} {PeerJ Computer Science}\ }\textbf {\bibinfo {volume} {10}},\ \bibinfo {pages} {e2245} (\bibinfo {year} {2024})}\BibitemShut {NoStop}%
\bibitem [{\citenamefont {{Le Cun}}\ \emph {et~al.}(1998)\citenamefont {{Le Cun}}, \citenamefont {Bottou}, \citenamefont {Orr},\ and\ \citenamefont {M{\"{u}}ller}}]{lecun-98x}%
  \BibitemOpen
  \bibfield  {author} {\bibinfo {author} {\bibfnamefont {Y.}~\bibnamefont {{Le Cun}}}, \bibinfo {author} {\bibfnamefont {L.}~\bibnamefont {Bottou}}, \bibinfo {author} {\bibfnamefont {G.~B.}\ \bibnamefont {Orr}},\ and\ \bibinfo {author} {\bibfnamefont {K.-R.}\ \bibnamefont {M{\"{u}}ller}},\ }in\ \href {http://leon.bottou.org/papers/lecun-98x} {\emph {\bibinfo {booktitle} {Neural Networks, Tricks of the Trade}}},\ \bibinfo {series and number} {Lecture Notes in Computer Science LNCS~1524}\ (\bibinfo  {publisher} {Springer Verlag},\ \bibinfo {year} {1998})\BibitemShut {NoStop}%
\bibitem [{\citenamefont {Agarap}(2019)}]{agarap2019deeplearningusingrectified}%
  \BibitemOpen
  \bibfield  {author} {\bibinfo {author} {\bibfnamefont {A.~F.}\ \bibnamefont {Agarap}},\ }\href {https://arxiv.org/abs/1803.08375} {\bibinfo {title} {Deep learning using rectified linear units (relu)}} (\bibinfo {year} {2019}),\ \Eprint {https://arxiv.org/abs/1803.08375} {arXiv:1803.08375 [cs.NE]} \BibitemShut {NoStop}%
\bibitem [{\citenamefont {Kingma}\ and\ \citenamefont {Ba}(2017)}]{kingma2017adammethodstochasticoptimization}%
  \BibitemOpen
  \bibfield  {author} {\bibinfo {author} {\bibfnamefont {D.~P.}\ \bibnamefont {Kingma}}\ and\ \bibinfo {author} {\bibfnamefont {J.}~\bibnamefont {Ba}},\ }\href {https://arxiv.org/abs/1412.6980} {\bibinfo {title} {Adam: A method for stochastic optimization}} (\bibinfo {year} {2017}),\ \Eprint {https://arxiv.org/abs/1412.6980} {arXiv:1412.6980 [cs.LG]} \BibitemShut {NoStop}%
\bibitem [{\citenamefont {Fey}\ and\ \citenamefont {Lenssen}(2019)}]{fey2019fast}%
  \BibitemOpen
  \bibfield  {author} {\bibinfo {author} {\bibfnamefont {M.}~\bibnamefont {Fey}}\ and\ \bibinfo {author} {\bibfnamefont {J.~E.}\ \bibnamefont {Lenssen}},\ }\href@noop {} {\bibinfo {title} {Fast graph representation learning with pytorch geometric}} (\bibinfo {year} {2019}),\ \Eprint {https://arxiv.org/abs/1903.02428} {arXiv:1903.02428 [cs.LG]} \BibitemShut {NoStop}%
\bibitem [{\citenamefont {Kipf}\ and\ \citenamefont {Welling}(2017)}]{kipf2017semisupervised}%
  \BibitemOpen
  \bibfield  {author} {\bibinfo {author} {\bibfnamefont {T.~N.}\ \bibnamefont {Kipf}}\ and\ \bibinfo {author} {\bibfnamefont {M.}~\bibnamefont {Welling}},\ }\href@noop {} {\bibinfo {title} {Semi-supervised classification with graph convolutional networks}} (\bibinfo {year} {2017}),\ \Eprint {https://arxiv.org/abs/1609.02907} {arXiv:1609.02907 [cs.LG]} \BibitemShut {NoStop}%
\bibitem [{\citenamefont {{Cranmer}}\ \emph {et~al.}(2020)\citenamefont {{Cranmer}}, \citenamefont {{Brehmer}},\ and\ \citenamefont {{Louppe}}}]{Cranmer_2020}%
  \BibitemOpen
  \bibfield  {author} {\bibinfo {author} {\bibfnamefont {K.}~\bibnamefont {{Cranmer}}}, \bibinfo {author} {\bibfnamefont {J.}~\bibnamefont {{Brehmer}}},\ and\ \bibinfo {author} {\bibfnamefont {G.}~\bibnamefont {{Louppe}}},\ }\href {https://doi.org/10.1073/pnas.1912789117} {\bibfield  {journal} {\bibinfo  {journal} {Proceedings of the National Academy of Science}\ }\textbf {\bibinfo {volume} {117}},\ \bibinfo {pages} {30055} (\bibinfo {year} {2020})},\ \Eprint {https://arxiv.org/abs/1911.01429} {arXiv:1911.01429 [stat.ML]} \BibitemShut {NoStop}%
\bibitem [{\citenamefont {Dalmasso}\ \emph {et~al.}(2024)\citenamefont {Dalmasso}, \citenamefont {Masserano}, \citenamefont {Zhao}, \citenamefont {Izbicki},\ and\ \citenamefont {Lee}}]{dalmasso2024likelihoodfreefrequentistinferencebridging}%
  \BibitemOpen
  \bibfield  {author} {\bibinfo {author} {\bibfnamefont {N.}~\bibnamefont {Dalmasso}}, \bibinfo {author} {\bibfnamefont {L.}~\bibnamefont {Masserano}}, \bibinfo {author} {\bibfnamefont {D.}~\bibnamefont {Zhao}}, \bibinfo {author} {\bibfnamefont {R.}~\bibnamefont {Izbicki}},\ and\ \bibinfo {author} {\bibfnamefont {A.~B.}\ \bibnamefont {Lee}},\ }\href {https://arxiv.org/abs/2107.03920} {\bibinfo {title} {Likelihood-free frequentist inference: Bridging classical statistics and machine learning for reliable simulator-based inference}} (\bibinfo {year} {2024}),\ \Eprint {https://arxiv.org/abs/2107.03920} {arXiv:2107.03920 [stat.ML]} \BibitemShut {NoStop}%
\bibitem [{\citenamefont {Green}\ \emph {et~al.}(2020)\citenamefont {Green}, \citenamefont {Simpson},\ and\ \citenamefont {Gair}}]{Green:2020hst}%
  \BibitemOpen
  \bibfield  {author} {\bibinfo {author} {\bibfnamefont {S.~R.}\ \bibnamefont {Green}}, \bibinfo {author} {\bibfnamefont {C.}~\bibnamefont {Simpson}},\ and\ \bibinfo {author} {\bibfnamefont {J.}~\bibnamefont {Gair}},\ }\href {https://doi.org/10.1103/PhysRevD.102.104057} {\bibfield  {journal} {\bibinfo  {journal} {Phys. Rev. D}\ }\textbf {\bibinfo {volume} {102}},\ \bibinfo {pages} {104057} (\bibinfo {year} {2020})},\ \Eprint {https://arxiv.org/abs/2002.07656} {arXiv:2002.07656 [astro-ph.IM]} \BibitemShut {NoStop}%
\bibitem [{\citenamefont {Dax}\ \emph {et~al.}(2021)\citenamefont {Dax}, \citenamefont {Green}, \citenamefont {Gair}, \citenamefont {Macke}, \citenamefont {Buonanno},\ and\ \citenamefont {Sch{\"o}lkopf}}]{Dax:2021tsq}%
  \BibitemOpen
  \bibfield  {author} {\bibinfo {author} {\bibfnamefont {M.}~\bibnamefont {Dax}}, \bibinfo {author} {\bibfnamefont {S.~R.}\ \bibnamefont {Green}}, \bibinfo {author} {\bibfnamefont {J.}~\bibnamefont {Gair}}, \bibinfo {author} {\bibfnamefont {J.~H.}\ \bibnamefont {Macke}}, \bibinfo {author} {\bibfnamefont {A.}~\bibnamefont {Buonanno}},\ and\ \bibinfo {author} {\bibfnamefont {B.}~\bibnamefont {Sch{\"o}lkopf}},\ }\href {https://doi.org/10.1103/PhysRevLett.127.241103} {\bibfield  {journal} {\bibinfo  {journal} {Phys. Rev. Lett.}\ }\textbf {\bibinfo {volume} {127}},\ \bibinfo {pages} {241103} (\bibinfo {year} {2021})},\ \Eprint {https://arxiv.org/abs/2106.12594} {arXiv:2106.12594 [gr-qc]} \BibitemShut {NoStop}%
\bibitem [{\citenamefont {Dax}\ \emph {et~al.}(2023)\citenamefont {Dax}, \citenamefont {Green}, \citenamefont {Gair}, \citenamefont {P{\"u}rrer}, \citenamefont {Wildberger}, \citenamefont {Macke}, \citenamefont {Buonanno},\ and\ \citenamefont {Sch{\"o}lkopf}}]{Dax:2022pxd}%
  \BibitemOpen
  \bibfield  {author} {\bibinfo {author} {\bibfnamefont {M.}~\bibnamefont {Dax}}, \bibinfo {author} {\bibfnamefont {S.~R.}\ \bibnamefont {Green}}, \bibinfo {author} {\bibfnamefont {J.}~\bibnamefont {Gair}}, \bibinfo {author} {\bibfnamefont {M.}~\bibnamefont {P{\"u}rrer}}, \bibinfo {author} {\bibfnamefont {J.}~\bibnamefont {Wildberger}}, \bibinfo {author} {\bibfnamefont {J.~H.}\ \bibnamefont {Macke}}, \bibinfo {author} {\bibfnamefont {A.}~\bibnamefont {Buonanno}},\ and\ \bibinfo {author} {\bibfnamefont {B.}~\bibnamefont {Sch{\"o}lkopf}},\ }\href {https://doi.org/10.1103/PhysRevLett.130.171403} {\bibfield  {journal} {\bibinfo  {journal} {Phys. Rev. Lett.}\ }\textbf {\bibinfo {volume} {130}},\ \bibinfo {pages} {171403} (\bibinfo {year} {2023})},\ \Eprint {https://arxiv.org/abs/2210.05686} {arXiv:2210.05686 [gr-qc]} \BibitemShut {NoStop}%
\bibitem [{\citenamefont {Dax}\ \emph {et~al.}(2025)\citenamefont {Dax}, \citenamefont {Green}, \citenamefont {Gair}, \citenamefont {Gupte}, \citenamefont {P{\"u}rrer}, \citenamefont {Raymond}, \citenamefont {Wildberger}, \citenamefont {Macke}, \citenamefont {Buonanno},\ and\ \citenamefont {Sch{\"o}lkopf}}]{Dax:2024mcn}%
  \BibitemOpen
  \bibfield  {author} {\bibinfo {author} {\bibfnamefont {M.}~\bibnamefont {Dax}}, \bibinfo {author} {\bibfnamefont {S.~R.}\ \bibnamefont {Green}}, \bibinfo {author} {\bibfnamefont {J.}~\bibnamefont {Gair}}, \bibinfo {author} {\bibfnamefont {N.}~\bibnamefont {Gupte}}, \bibinfo {author} {\bibfnamefont {M.}~\bibnamefont {P{\"u}rrer}}, \bibinfo {author} {\bibfnamefont {V.}~\bibnamefont {Raymond}}, \bibinfo {author} {\bibfnamefont {J.}~\bibnamefont {Wildberger}}, \bibinfo {author} {\bibfnamefont {J.~H.}\ \bibnamefont {Macke}}, \bibinfo {author} {\bibfnamefont {A.}~\bibnamefont {Buonanno}},\ and\ \bibinfo {author} {\bibfnamefont {B.}~\bibnamefont {Sch{\"o}lkopf}},\ }\href {https://doi.org/10.1038/s41586-025-08593-z} {\bibfield  {journal} {\bibinfo  {journal} {Nature}\ }\textbf {\bibinfo {volume} {639}},\ \bibinfo {pages} {49} (\bibinfo {year} {2025})},\ \Eprint {https://arxiv.org/abs/2407.09602} {arXiv:2407.09602 [gr-qc]} \BibitemShut {NoStop}%
\bibitem [{\citenamefont {Alvey}\ \emph {et~al.}(2024)\citenamefont {Alvey}, \citenamefont {Bhardwaj}, \citenamefont {Domcke}, \citenamefont {Pieroni},\ and\ \citenamefont {Weniger}}]{Alvey:2023npw}%
  \BibitemOpen
  \bibfield  {author} {\bibinfo {author} {\bibfnamefont {J.}~\bibnamefont {Alvey}}, \bibinfo {author} {\bibfnamefont {U.}~\bibnamefont {Bhardwaj}}, \bibinfo {author} {\bibfnamefont {V.}~\bibnamefont {Domcke}}, \bibinfo {author} {\bibfnamefont {M.}~\bibnamefont {Pieroni}},\ and\ \bibinfo {author} {\bibfnamefont {C.}~\bibnamefont {Weniger}},\ }\href {https://doi.org/10.1103/PhysRevD.109.083008} {\bibfield  {journal} {\bibinfo  {journal} {Phys. Rev. D}\ }\textbf {\bibinfo {volume} {109}},\ \bibinfo {pages} {083008} (\bibinfo {year} {2024})},\ \Eprint {https://arxiv.org/abs/2309.07954} {arXiv:2309.07954 [gr-qc]} \BibitemShut {NoStop}%
\bibitem [{\citenamefont {Dimitriou}\ \emph {et~al.}(2024)\citenamefont {Dimitriou}, \citenamefont {Figueroa},\ and\ \citenamefont {Zaldívar}}]{Dimitriou_2024}%
  \BibitemOpen
  \bibfield  {author} {\bibinfo {author} {\bibfnamefont {A.}~\bibnamefont {Dimitriou}}, \bibinfo {author} {\bibfnamefont {D.~G.}\ \bibnamefont {Figueroa}},\ and\ \bibinfo {author} {\bibfnamefont {B.}~\bibnamefont {Zaldívar}},\ }\href {https://doi.org/10.1088/1475-7516/2024/09/032} {\bibfield  {journal} {\bibinfo  {journal} {Journal of Cosmology and Astroparticle Physics}\ }\textbf {\bibinfo {volume} {2024}}\bibinfo  {number} { (09)},\ \bibinfo {pages} {032}}\BibitemShut {NoStop}%
\end{thebibliography}%

\end{document}